\documentclass{article}
\usepackage[english]{babel}

\usepackage[letterpaper,top=2cm,bottom=2cm,left=3cm,right=3cm,marginparwidth=1.75cm]{geometry}

\usepackage{amsmath}
\usepackage[colorlinks=true, allcolors=blue]{hyperref}
\usepackage{authblk}
\usepackage{graphicx}
\usepackage[utf8]{inputenc}
\usepackage[english]{babel}
\usepackage{amsfonts}
\usepackage{multirow}
\usepackage{amssymb}
\usepackage[usenames, dvipsnames]{color}
\usepackage{hyperref}
\usepackage[capitalise]{cleveref}
\usepackage{enumitem}
\usepackage[nice]{nicefrac}
\usepackage{graphicx}  
\usepackage{lscape}
\usepackage{enumitem}
\usepackage{comment}
\newcommand\redsout{\bgroup\markoverwith{\textcolor{red}{\rule[0.5ex]{2pt}{0.4pt}}}\ULon}
\newcommand\bluesout{\bgroup\markoverwith{\textcolor{blue}{\rule[0.5ex]{2pt}{0.4pt}}}\ULon}

\usepackage{csquotes}
\usepackage[style=nature]{biblatex}
\newcommand{\ADhide}[1]{{}}
\newcommand{\SMhide}[1]{{}}

\title{Half-quantized Hall Plateaus in the Confined Geometry of Graphene}
\author[1,2]{Preeti Pandey$^*$}
\author[3]{Sourav Manna$^*$}
\author[1,2]{Kristiana N. Frei$^*$}
\author[1,2]{Jerin Saji}
\author[4]{Anne Denis}
\author[2]{Alexander Savin}
\author[5]{Kenji Watanabe}
\author[5]{Takashi Taniguchi}
\author[1,2]{Pertti J. Hakonen}
\author[6,3]{Ankur Das}
\author[1,2]{Manohar Kumar$^\#$}
\affil[1]{Department of Applied Physics, School of Science, Aalto University, P.O. Box 15100, FI-00076, Finland}
\affil[2]{QTF Centre of Excellence, Department of Applied Physics, Aalto University, P.O. Box 15100, FI-00076 Aalto, Finland}
\affil[3]{Department of Condensed Matter Physics, Weizmann Institute of Science, Rehovot 7610001, Israel}
\affil[4]{Laboratoire de Physique de l’Ecole Normale Supérieure, ENS, Université PSL, CNRS, Sorbonne Université, Université Paris Cité, F-75005 Paris, France}
\affil[5]{National Institute for Materials Science, 1-1 Namiki, Tsukuba 305-0044, Japan}
\affil[6]{Department of Physics, Indian Institute of Science Education and Research (IISER) Tirupati, Tirupati 517619, India}
\affil[$*$]{These authors contributed equally to this work}
\affil[$\#$]{Corresponding author: manohar.kumar@aalto.fi}

\begin{document}
\maketitle
\begin{abstract}
Since the ground-breaking discovery of the quantum Hall effect, half-quantized quantum Hall plateaus have been some of the most studied and sought-after states.  Their importance stems not only from the fact that they transcend the composite fermion framework used to explain fractional quantum Hall states (such as Laughlin states). Crucially, they hold promise for hosting non-Abelian excitations, which are essential for developing topological qubits — key components for fault-tolerant quantum computing.
In this work, we show that these coveted half-quantized plateaus can appear in more than one unexpected way.
We report the observation of fractional states with conductance quantization at $\nu_H = 5/2$ arising due to charge equilibration in the confined region of a quantum point contact in monolayer graphene.
\end{abstract}

\section*{Introduction}
Anyons are exotic quasiparticles that exist in certain two-dimensional systems, such as those exhibiting the fractional quantum Hall (FQH) effect \cite{PhysRevLett.48.1559, PhysRevLett.50.1395}. Compared to ordinary bosons or fermions, anyons have unique properties. They carry fractional values of the electric charge $e$ and exchange statistics quantified by the statistical phase $\theta$ arising in the phase factor $e^{i\theta}$ acquired upon the exchange of two particles \cite{Leinaas1977, PhysRevLett.49.957, PhysRevLett.48.1144}. Unlike for bosons or fermions, $\theta$ is not limited to 0 or $\pi$.  
For Laughlin states, which describe many FQH states, this phase is typically $2\pi/q$, where $q$ is the denominator of the filling fraction $\nu = p/q$ with $p$ and $q$ both odd numbers. For example, $\nu=1/3$ has a fractional $e/3$ charge and a statistical phase of $2\pi/3$ \cite{PhysRevLett.50.1395}.

The fractional charge of the anyons was already measured almost three decades ago via shot noise \cite{Mahalu1997, PhysRevLett.79.2526}. Recently, their fractional statistics have been observed in collider experiments \cite{Bartolomei2020, PhysRevLett.116.156802} and in interferometer experiments, including Fabry-P\'erot \cite{Nakamura2020, PhysRevX.13.041012} and Mach-Zehnder interferometers \cite{Kundu2023}. 

\par Contrary to the odd-denominator fractional quantum Hall (FQH) states, which are fairly well understood in terms of Laughlin wavefunctions, the physics for even-denominator FQH states is more involved \cite{PhysRevLett.50.1395, PhysRevLett.63.199}. These even denominator FQH states are observed in high-quality two-dimensional electron gases (2-DEGs) \cite{Chung2021}.
The first observation of such a state, having a filling fraction of $\nu=5/2$ \cite{Willet1987}, is widely accepted to carry paired composite fermions carrying non-Abelian quantum statistics in its ground state \cite{PhysRevLett.80.1505}.
These quasiparticle excitations are expected to facilitate fault-tolerant topological quantum computations \cite{RevModPhys.80.1083}. Although agreed upon that the ground state is non-Abelian, its exact nature is still debated.
The numerical studies support Pfaffian and anti-Pfaffian states \cite{PhysRevLett.80.1505, PhysRevLett.106.116801}, which have identical Hall conductances but very different edge constituent modes.
This leads to the possibility of distinguishing them via thermal conductivity measurements, however, previous such experiments point to particle-hole symmetric Pfaffian states \cite{Banerjee2018, Dutta2022, Dutta2022_Iso, paul2024}.

Graphene is an interesting material for studying these exotic FQH states. Contrary to conventional 2-DEGs, it is single-atom-thick. It thus has stronger electron-electron interactions leading it to host a series of unconventional FQH states \cite{Kumar2018} and strong quantum corrections in conductivity of the Fermi sea of composite fermions \cite{LaitinenHalfFilling}. Even-denominator FQH states have been reported in the $N = 1$ Landau level in bilayer graphene \cite{Ki2014}. Monolayer graphene displayed half-quantization at higher Landau levels \cite{Zibrov2018, Kim2019}, supporting 221 parton states for their origin.
In that regard, graphene systems may emerge as one of the promising platforms for probing the poorly understood $5/2$ plateau.
\\

\begin{figure}[ht!]
    \centering
    \includegraphics[width=0.9\columnwidth]{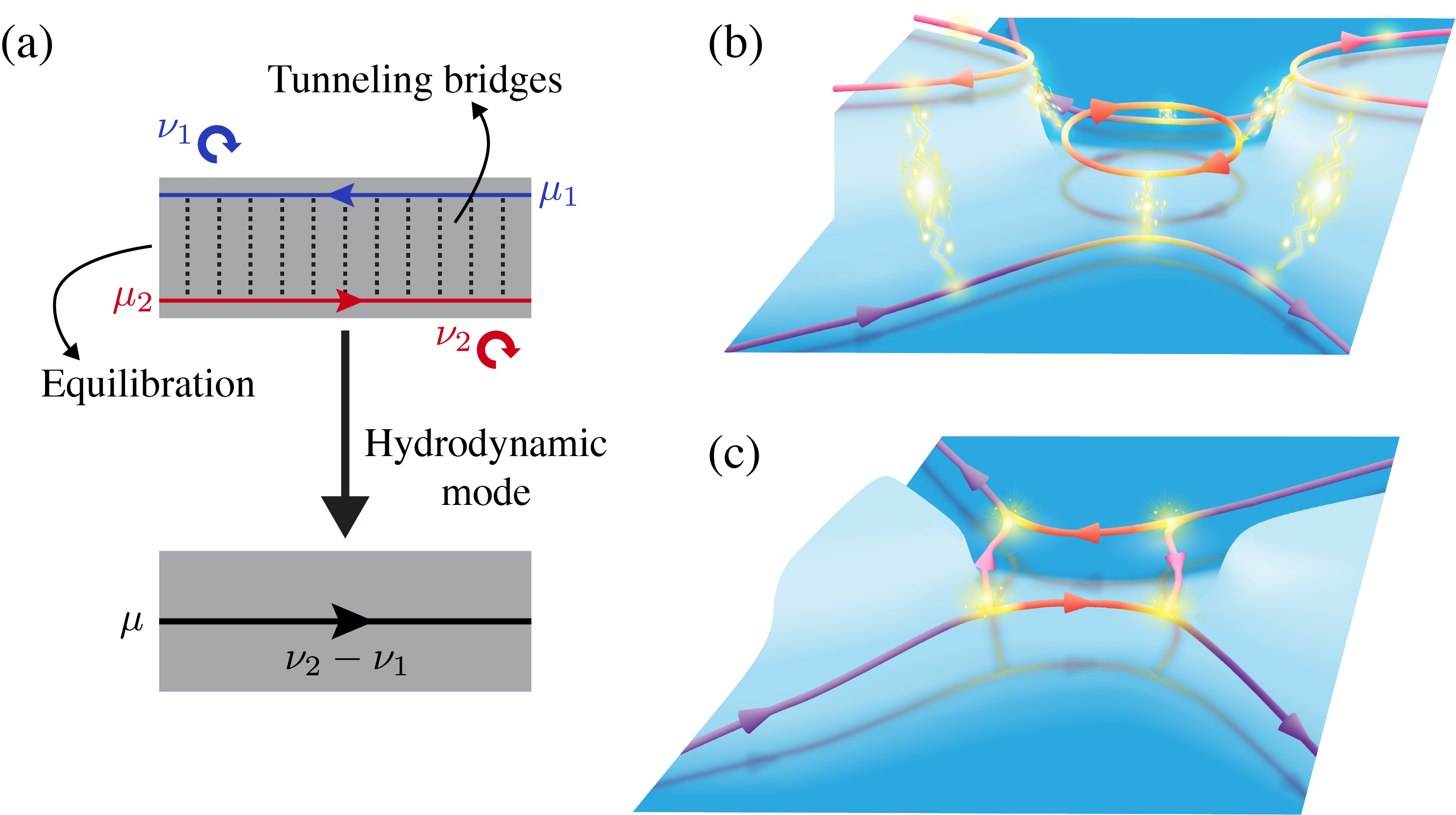}
    \caption{(a) Schematic view of the equilibrated edge modes at fractional fillings. Charge equilibration through tunneling bridges between two counter-propagating modes with filling $\nu_1$ and $\nu_2 (>\nu_1)$ gives rise to a hydrodynamic mode with an effective conductance of $\nu_2-\nu_1$. Circular arrows show the chirality of charge propagation of the corresponding filling fractions. Here $\mu_1$ ($\mu_2$) denotes the initial chemical potential for the mode with filling $\nu_1$ ($\nu_2$) and $\mu$ is the the chemical potential of the equilibrated mode. (b)-(c) Illustrative view of the equilibration process at the QPC. The chemical potential under the split top gates can be tuned away from the chemical potential in the bulk. This gives rise to a chemical potential profile that can host edge states with different fractional fillings in three regions simultaneously; inside the QPC, under the top gates, and in the bulk (b). These equilibrate via tunneling, which alters the chemical potential profile, and the effective description is merged edge states in a hydrodynamic regime (c).}
    \label{fig:cartoon}
\end{figure}

\par Recently, it was shown that the confinement potential from a quantum point contact (QPC) stabilizes the $\nu = 3/2$ and $\nu = 1/2$ plateaus emerging from the bulk filling of $\nu = 5/3$ and $\nu = 2/3$ \cite{Yan2023, Fauzi2023, Nakamura2023}. The concept of edge equilibration was introduced \cite{Banerjee2017, Srivastav2021, Protopopov2017, PhysRevLett.132.136502, SMshort, SMlong, SMAD} to explain these anomalous behaviors. Here, the idea is that several chiral edge states will become interwoven. Eventually, corresponding modes will equilibrate due to local tunnelling, allowing the different modes to exchange heat and charge. We can understand this by introducing two different thermodynamic quantities to be defined locally: chemical potential $\mu$ and temperature. Now, these two can reach a steady state (equilibrate) independent of each other, provided the system length is large enough. When this happens, the edge looses its coherence and becomes an effective system in the hydrodynamic regime. The equilibration of the chemical potential (also known as charge equilibration) can lead to new and peculiar behaviours, see \cref{fig:cartoon}. It can produce unexpected plateaus \cite{SMshort, SMlong, SMAD, Nakamura2023} and help us to distinguish the different candidate states of stabilized plateaus \cite{PhysRevLett.132.136502}. These processes are more effective in confined regions \cite{Christian2020, Fauzi2023}. While the charge is fully equilibrated, the electrical conductance becomes insensitive to edge reconstruction.

\par To define the confined regions, the local modulation of the charge density to zero is a prerequisite, for which the voltage-controlled constriction defined by a top gate is a commonly used tool. Unlike in GaAs/AlGaAs 2-DEGs, the absence of the band gap in monolayer graphene makes the design of the QPC a non-trivial technical feat. Here, tweaking the chemical potential on the top gate will modulate the charge density either in the $p$-doped or $n$-doped region forming a $npn$ or $pnp$ junction \cite{Young2009, Stander2009}. However, at high magnetic fields, Landau levels are formed which have a large gap energy, allowing the top-gated structure to act like a QPC. In recent years, several approaches have been used to design QPCs in graphene \cite{Williams2007, Zimmermann2017, Kun2020}. Out of these constructions, the split-gate QPC is the most promising one for controlling the transmission of the quantum Hall edge states \cite{Cohen2023}. 
 
\par Quantum Hall transport measurements in the two-terminal geometry of graphene $pn~(np)$ or $pnp~(npn)$ junctions have already been performed, where unusual quantization at fractional values of $e^2/h$, $h$ being the Planck's constant, has been observed \cite{Ozyilmaz2007, Williams2007}. The quantization of the filling fractions of $\nu = 3/2$ and $5/2$ in diagonal conductance is reported in earlier works on a junction \cite{Amet2014} and a device with a QPC \cite{Zimmermann2017}. Also, graphene Hall bar devices with two symmetric and independent back gates forming \textit{pn} junction have shown the formation of  $\nu = 3/2$ and $5/2$ in longitudinal resistance measurements across the $pn$ junction \cite{Nikolai2015}. The origin of these states is related to the spin-selective equilibration of quantum Hall edge channels (detailed in Supplementary \cref{Note4a}). 

\par In this work, we show how unexpected half-quantized quantum Hall plateaus can appear in a confined geometry. We focus on the state with $\nu_H = 5/2$, which we observed as a stabilization of a plateau in the Hall resistance $R_H = h/(\nu_H e^2)$. The quantization of this state in the $N = 1$ Landau level of graphene is an intriguing case. It is equivalent to $\nu = 1/2$ in GaAs/AlGaAs 2-DEGs, which is a compressible state and thus expected to be compressible in graphene as well. Contrary to this, we observed quantization at $\nu_H = \pm 5/2$ in two distinct regimes: 
a) transport across FQH fluid sections with different filling factors (``Ordinary"), and b) transport across an FQH fluid point contact bridging between two special Fermi-liquid-like reservoirs facilitated by graphene's 0th Landau level (``Out-of-Ordinary"). These plateaus appear for reasons specific to these regimes that we will describe in detail in this article. Generic to both cases, the different quantum Hall plateaus are stabilized due to non-equilibrium inter-edge interactions and tunnelling in the confined geometry of the QPC.

\section*{Results}
\subsection*{Graphene Hall bar with split-gate QPC}
\begin{figure}[htbp]
    \centering
    \includegraphics[scale=1]{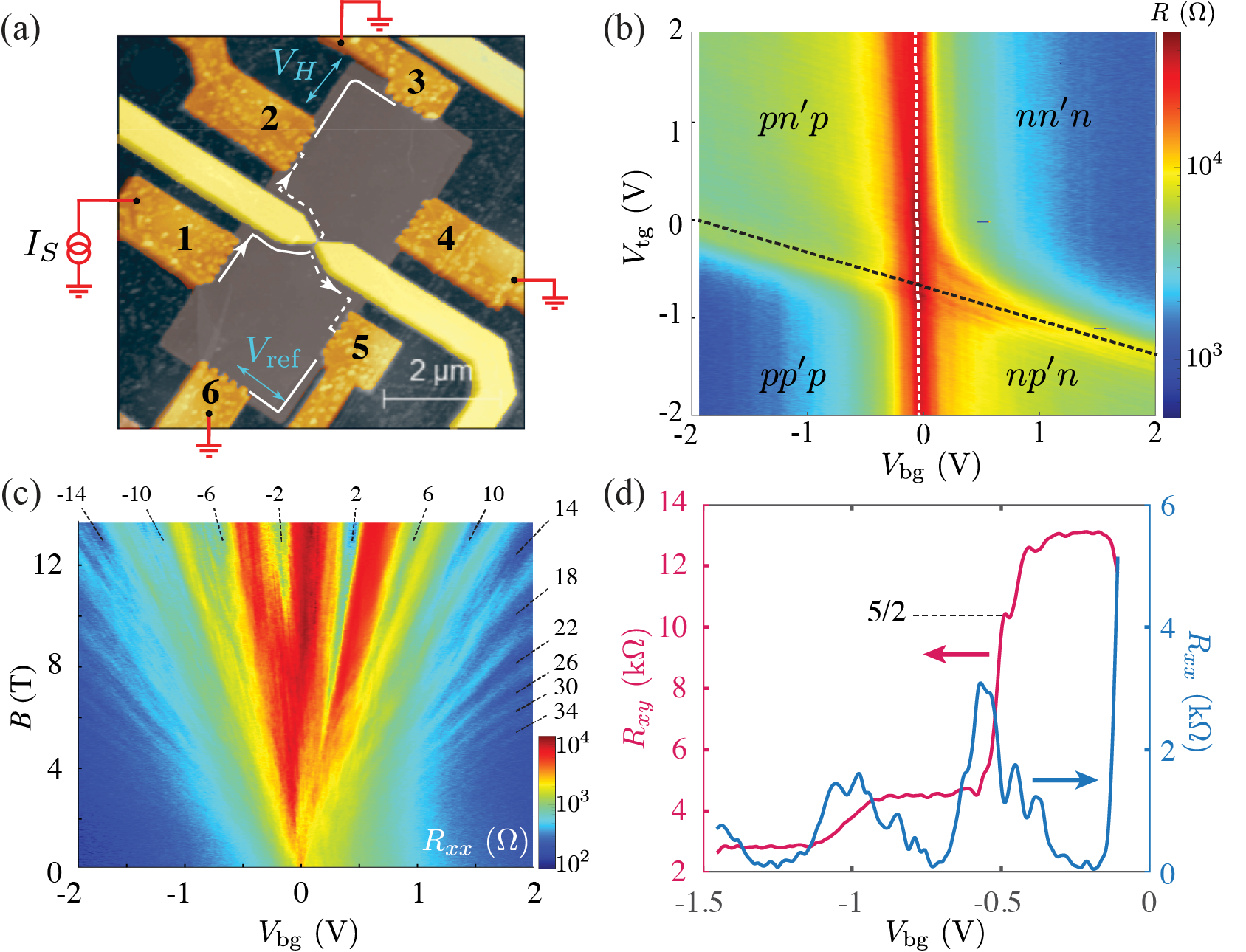}
    \caption{ (a) Atomic scale microscopy image of graphene Hall bar sample  (false color): graphene is shaded in brown color, ohmic edge contacts are marked with bronze color, and the local top gate (QPC) with light yellow. In the depicted configuration, current $I_S$ is applied to contact 1 while contacts 3, 4, and 6 are grounded, such that the signal transmitted through the QPC $R_H$ is measured by voltage $V_H$ between contacts 2 and 3. Similarly, the reflected signal $R_{\text{ref}}$ is identified by $V_\text{ref}$ across contacts 5 and 6. The configuration for measuring $R_{xx}$ and $R_{xy}$ are detailed in Methods. (b) Gate characteristics for Hall bar sample at zero magnetic field. The white dashed line marks the Dirac peak in the back gate $V_\text{bg}$ sweep and the black dashed line marks the resistance maximum with top gate $V_\text{tg}$ sweeping. The slope of this line is used to calculate the top to back gate coupling constant $m_\text{tg}$. (c) $R_{xx}$ measured as a function of $V_\text{bg}$ and magnetic field $B$. Quantum Hall steps are seen as dips in $R_{xx}$ when $|\nu_H| \simeq 2, 6, 10, 14, 18, 22, 26, 30$ and 34. The corresponding steps in Hall resistance $R_{xy}$ are shown in Supplementary \cref{HallR}. Note that only the dominant Hall steps are marked here. (d) The line trace of $R_{xy}$ and $R_{xx}$ measured at $B=-12.5$ T and $T = 20$ mK while keeping the QPC at the charge neutrality point $V_\text{tg} = -0.75$ V. The dominant fractional quantization is at $\nu_H = 2+1/2$.}
    \label{Fig1}
\end{figure}

Hall bar samples with six ohmic contacts were fabricated on hexagonal boron nitride encapsulated graphene heterostructures with graphite back gates (see \cref{Fig1}a). A metallic split-top gate is symmetrically placed between the two innermost voltage contacts to form a QPC. We measured four Hall bar samples to confirm the reproducibility of the results and limit ourselves to the presentation of data from two samples in this report. See Methods for details on the fabrication as well as the classification of the samples.
The coupling between the top gate and the bottom back gate is characterized by sweeping the back gate while keeping the top gate at a constant voltage. The two-gate sweep is shown in \cref{Fig1}b. The Dirac peak is close to zero for the back gate charge density modulation and slightly negative for the top gate.
The top split-gate QPC and back gate are strongly coupled with coupling constant $m_\text{tg} = -0.36$ calculated from the slope of the black dashed line. We note that earlier reports on similar samples show weaker coupling of the QPC and back gate potential \cite{Zimmermann2017}. An asymptotic modulation of the charge density can be identified in the unipolar region of the $nn^\prime n$ region, which is missing in the bipolar region of the $pn^\prime n$ or $np^\prime p$ regions due to the shallow potential gradient at the junctions. Potential mapping is modelled using COMSOL Multiphysics\textsuperscript \textregistered finite element analysis of the electrostatic field across the split gate; 
see Supplementary \cref{Note3}.
As an indicator of the confinement by the split gate, conductance quantization steps at 14$e^2/h$, 16$e^2/h$, and 24$e^2/h$ at zero magnetic fields were seen when the back gate was modulated while keeping the top gate at the Dirac point.
\par Upon the application of a magnetic field, measurements are done in two configurations: first, the sample is characterized by measuring $R_{xx}$ and $R_{xy}$, and to quantify the effect of the QPC we measured the transmitted and reflected current for probing $R_{H}$ and $R_{ref}$, respectively (see Methods). The usual sequence of resistance quantization in steps of $h/(4e^2)$ reflecting the 4-fold spin-valley degeneracy (2 for spin and 2 for valley) of each Landau level is observed in $R_{xy}$ (see Supplementary \cref{HallR}). In $R_{xx}$, the most prominent observed quantum Hall states $\nu_H \equiv \pm 4(N+1/2) = 6, 10, 14$ and 18 where $N$ is the Landau Level index, are  fully resolved at $B>3$ T, shown in \cref{Fig2}c. The $\nu_H = 2$ state is fully resolved at $B > 6$ T. Odd-numbered filling factor states with $\nu_H = 3, 5, 7...$ are observed, but less robust than the degenerate states, indicating partial spin and valley polarization. At the Dirac point, a resistive peak is observed in the longitudinal resistance (see \cref{Fig1}c) and the transverse resistance traverses through a low resistance value (see Supplementary \cref{HallR}), indicating full mixing of hole and particle Hall states. This feature of interchanging bulk and edge transport at the charge neutrality point is consistent with earlier reports \cite{Abanin2007, Joseph2008}.
At higher magnetic fields, $|B| = 12.5$ T, in Landau level $N=1$, a plateau at $\nu_H = 5/2$ is observed in $R_{xy}$ along with a dip in $R_{xx}$, see \cref{Fig1}d. This observation indicates the formation of incompressible states at $\nu_H = 5/2$ in contradiction to the much-expected compressible states in graphene. We note that recent experiments observe half-quantization exclusively in higher Landau levels of graphene \cite{Kim2019}. To understand the origin of this unconventional plateau, we performed detailed studies on the magneto-resistance of the Hall bar samples in two field regimes $|B| \sim 10$ T and $|B| \sim 13$ T.
                    
\subsection*{Modulation of quantum Hall transport and quantized plateau at $\nu_H = 5/2$}

 \begin{figure}[ht!]
    \centering
    \includegraphics[scale=1]{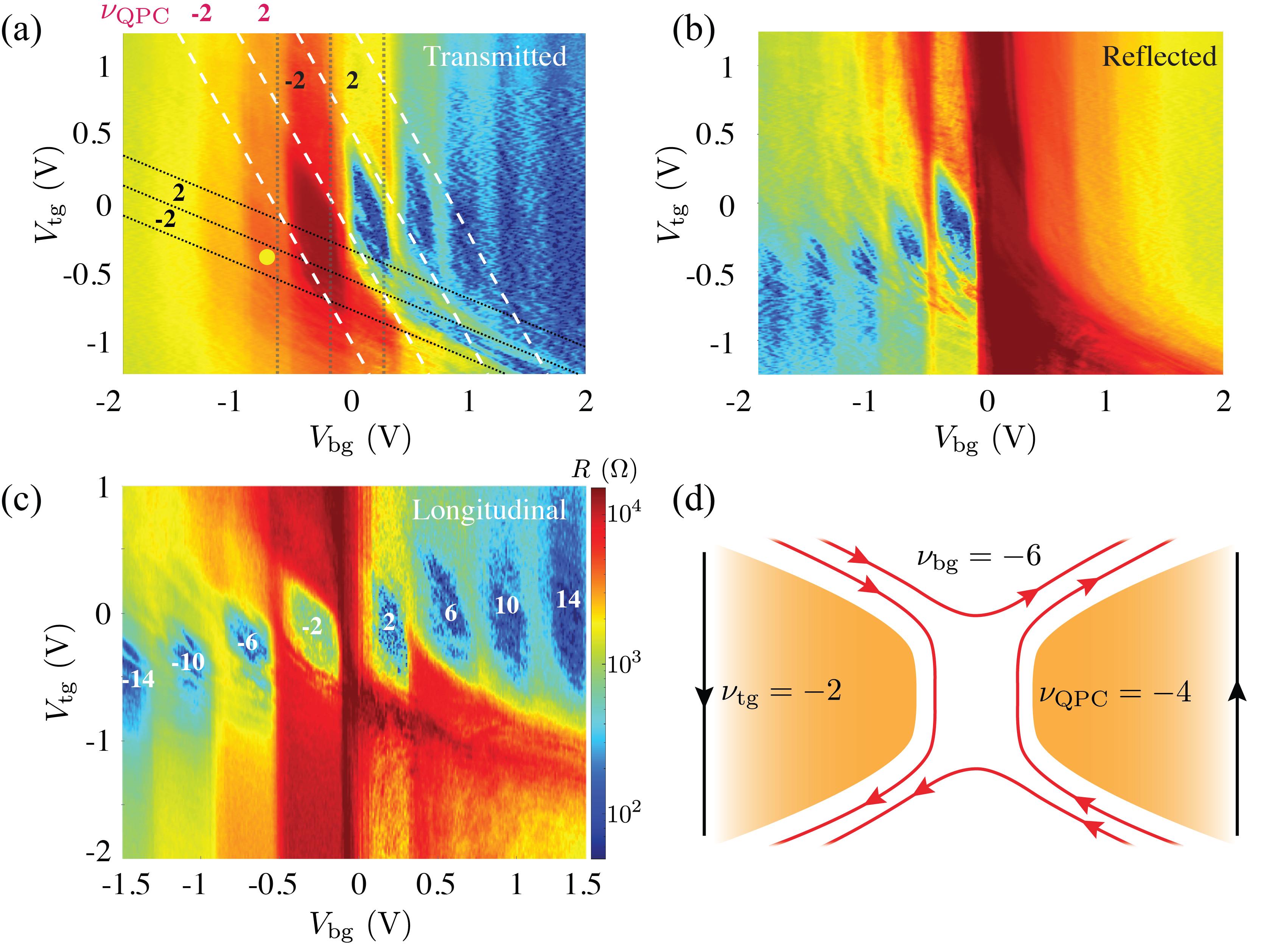}
    \caption{Quantum point controlled quantum transport in Hall bar sample at $10$ T: (a) transmitted $R_\text{H}$, (b) reflected $R_\text{ref}$, and (c) longitudinal $R_{xx}$. The color bar in (c) is shared between the figures. The grey dotted lines marked in (a) indicate the constant filling factor $\nu_\text{bg}$ in the bulk of the graphene due to the back gate. The black dotted lines are estimated fillings $\nu_\text{tg}$ under the top gate. The white dashed lines denote the filling $\nu_\text{QPC}$ in the region between the split gates of the QPC. In (c), the parallelogram structures show low resistance regimes that form when $\nu_{\text{bg}} \simeq \nu_{\text{QPC}}$. The filling for each parallelogram is marked in the figure. The parallelograms for $\pm 2$ have slightly larger $R_{xx}$, showing the finite interaction of the edge states in the QPC region. In the absence of any inter-edge interaction, \textit{i.e.} equilibration processes, each parallelogram should show minimal resistance. (d) A schematic of degenerate edge-state transport across the QPC. The effective filling factors for different regions are shown for the case marked by the yellow dot in (a). The Supplementary \cref{Note4a} details the edge-state equilibration model.
    }
    \label{Fig2}
\end{figure}

The transverse resistance $R_{xy}$ and longitudinal resistance $R_{xx}$ are fully resolved and robust at the high magnetic fields. The transport spectroscopy of the Hall bar sample in the measurement configuration as depicted in \cref{Fig1}a reveals the modulation of the Hall transport across the split-gate QPC probing $R_H$ and $R_\text{ref}$. \cref{Fig2}a and b show $R_H$ and $R_\text{ref}$ measured at $B = 10$ T. Here vertical strips in the color plot mark the $\nu_{\text{bg}} =\pm 2, 6, 10, 14$ and 18 filling factors due to the back gate modulation. There is a clear distinction between varying the charge density from $p$ to $n$: on a $p$-type graphene Hall bar, the electronic current traverses across the QPC, and Hall voltage $V_H$ is measured across the junction between contacts 2 and 3. With a reversal of carriers to $n$-type, the current from contact 1 sinks to the cold ground at contact 6, thus dominating the low resistance region. The quantum Hall regime shows regions with the lowest resistance in the order of 10 $\Omega$. The presence of the QPC further affects the transport. The diagonal black dashed line with slope $m_\text{tg}$ (see \cref{Fig1}b) gives estimated filling factors  $\nu_{\text{tg}}$ under the top gate. The asymptotic modulation of the charge density shows the coupling of the top and bottom gates (see Supplementary \cref{Note4}). In our sample, the back gate contribution mostly dominates the charge density, except for close to the top gate Dirac peak. 

Additionally, distinct parallelograms are seen in the transport spectroscopy where the slope of the edge of the parallelogram $m_\text{QPC} = -1.55$ is different from $m_\text{tg}$ that defines $\nu_{\text{tg}}$. These two distinct slopes are attributed to the top-to-back gate coupling and show that the narrow region between the split gates has a different charge density compared to the bulk and under the top gates. This region has filling factor $\nu_{\text{QPC}}$, which is strongly coupled to the back gate. A schematic detailing the edge-state transport with corresponding filling factors is shown in \cref{Fig2}d. When $\nu_{\text{bg}} \simeq \nu_{\text{QPC}}$, all of the forward currents should pass through the QPC, which forms parallelograms of low resistance. In \cref{Fig2}c, these can be seen for both positive and negative back gate voltages $V_\text{bg}$ in the longitudinal resistance $R_{xx}$. Interestingly, a slightly higher resistance is observed in the parallelograms at $\nu_\text{bg}\simeq\nu_\text{QPC}\simeq \pm2$. This hints at finite interaction taking place between the edge states in the vicinity of the QPC, allowing for unconventional fractions values such as $\nu_H= 5/2$ to arise. To make this more apparent, we will study the evolution of the $5/2$ plateau under the influence of perturbation due to different external parameters.

\begin{figure}[htbp]
    \centering
    \includegraphics[]{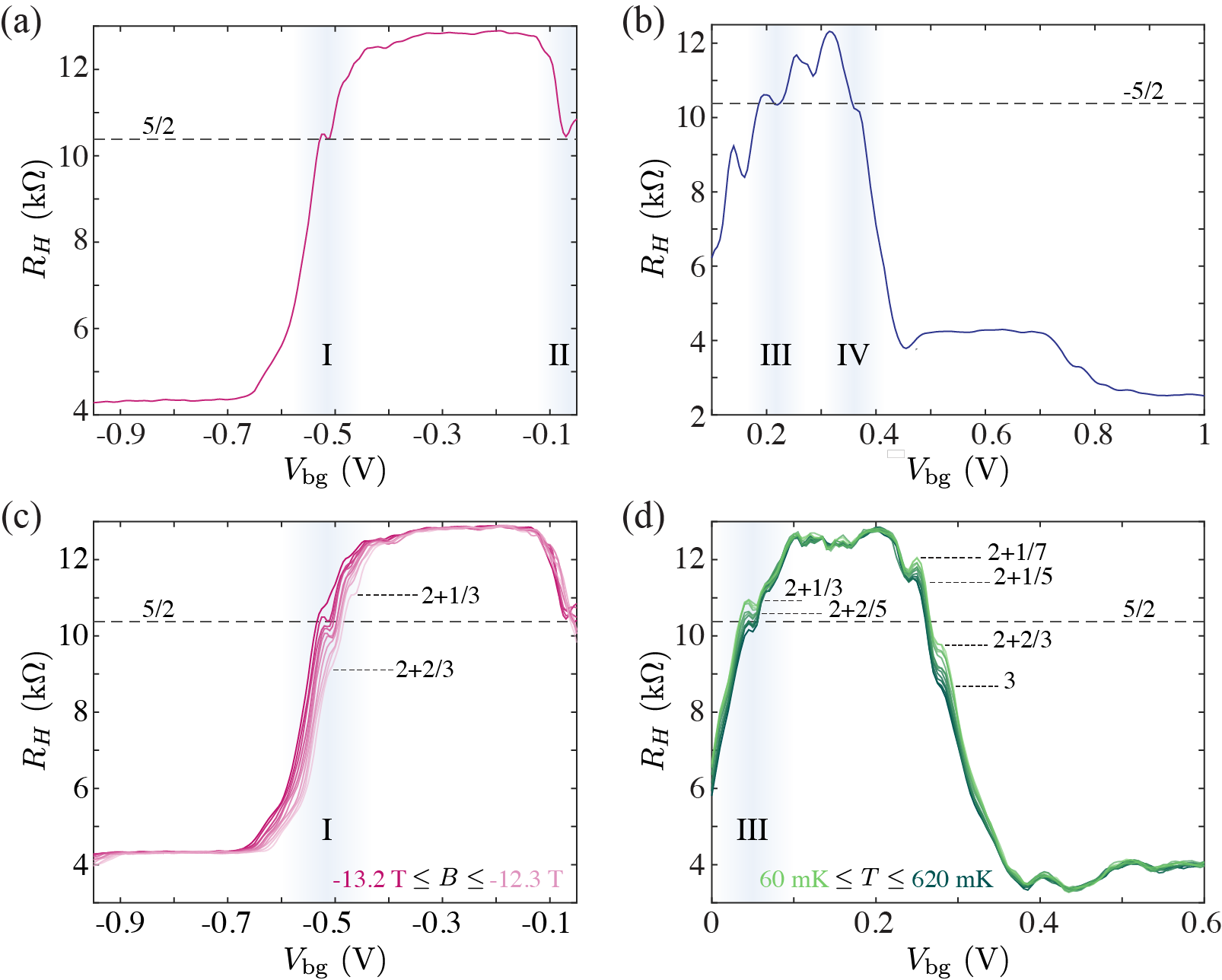}
    \caption{
    Four distinct regions (I-IV) for the occurrence of half-quantization in the Landau level N=1 are observed in $R_H$ as a function of $V_\text{bg}$. The QPC is biased to $V_\text{tg} = -0.75$ V and $B = -13.1$ T in (a) and $B = -9.75$ T in (b). The quantization of the half-filled state $\nu_H = \pm 5/2$ can be seen in four regions, I (``Ordinary") and II (``Out-of-Ordinary") for $V_\text{bg} < 0$ and III (``Out-of-Ordinary") and IV (``Ordinary") for $V_\text{bg} > 0$. (c) The Hall plateau shown in (a) is measured with a change in the magnetic field in steps of $100$ mT, ranging from $B=-13.2$ T (dark) to $B=-12.3$ T (light). (d) The temperature stability of the half-quantization is probed for region III at $B = 10$ T and $V_\text{tg} = 0$ V. The 5/2 step is absent in region IV here because the QPC is kept at zero voltage. $T$ ranges from $60$ mK (light) to $620$ mK (dark) in steps of 50 mK. With decreasing temperature the plateau shifts towards lower filling, which can also be seen for steps at other fractional values. The $\nu_H = 5/2$ state is stable around 410 to 560 mK and upon decreasing $T$ shifts to $\nu_H=2+2/5$ and finally $\nu_H=2+1/3$ at 60 mK.}
    \label{Fig3}
\end{figure}

\par In \cref{Fig3}a and b, we show the typical transport measurement of our Hall bar sample while sweeping the back gate voltage at a fixed magnetic field ($B= -13.10$ T and $B = -9.75$ T, respectively) and the top gate voltage tuned to the Dirac peak. A stark contrast can be seen; the $\nu_H = 2$ plateau is not fully developed in \cref{Fig3}b due to the lower magnetic field. Here we also see oscillations in $R_H$ at low $V_\text{bg}$ values. These flatten out into plateaus at higher temperatures and with the variation of the gate potential. To our understanding, these features are related to localized states close to the QPC. Due to non-linear screening of the potential fluctuations in the QPC region, many localized states or a network of compressible/incompressible strips is formed \cite{Baer2014}. The coupling of these states with each other and with the bulk edge states results in finite-size resonant structures in the measured Hall voltages.

\par The traces in \cref{Fig3}a-d show the quantization of $|\nu_H|= 5/2$ in four regions that can be attributed to either the ``Ordinary" (regions I and IV) or ``Out-of-Ordinary" (II and III) formation of the state. The ``Ordinary" regions form the plateau due to the equilibration of different edge states at the QPC, which while named ``Ordinary", is an unusual observation for given filling. The ``Out-of-Ordinary" regions show the plateaus in an even more unexpected scenario, specifically when equilibration takes place with a metallic bulk that is observed around the center of the 0th Landau level. We studied the evolution of the half-quantization in these regions with respect to the gate voltages, magnetic field, and temperature. It became evident that these plateaus are very sensitive to the details of the parameter space. Perturbation generally causes the half-quantized states to break down and stabilize at nearby conventional fractional fillings. \cref{Fig3}c shows how $\nu_H = 5/2$ in region I measured at $B = -13.1$ T (\cref{Fig3}a) gradually shifts to $\nu_H = 2+2/3$ when the magnetic field is lowered to $B = -12.3$ T.  A similar trend is observed for region IV in \cref{Fig3}b, shown in Supplementary \cref{sevenby2evolve}b.  
The temperature dependence of the $\nu_H=5/2$ plateau in region III is seen in \cref{Fig3}d. For applied parameters $B= 10$ T and $V_\text{tg}= 0$ V, the plateau is stable from $T=$ 410 mK to $T=$ 560 mK. Lowering the temperature evolves the plateau to $\nu_H = 2+2/5$ and finally $\nu_H = 2+1/3$ at $T = 60$ mK. More studies, including the top gate dependence, are detailed in Supplementary \cref{NoteStability}. As we present in the equilibration model, interactions between the bulk edge states $\nu_{\text{bg}}$ and the localized states $\nu_{\text{QPC}}$ result in a modification of the Hall filling factors $\nu_H$. With a change in perturbation parameters, the local states within the QPC favor different filling fractions, depending on their gap energy or the degree of coupling among other localized states and with the bulk edge channels. We also note that all four regions are not seen simultaneously in a single sweep of $V_\text{bg}$ for a set of fixed parameters. Quantization of steps at other half fillings is shown in Supplementary \cref{sevenby2evolve}.

\subsection*{Theoretical explanation for the origin of the $5/2$-plateaus}

The four regions exhibiting half-quantization at $\nu_H = 5/2$ can be quantified in terms of the interplay among $\nu_{\text{tg}}$, $\nu_{\text{QPC}}$ and $\nu_{\text{bg}}$ as follows

        \begin{enumerate}[label=\Roman*.]
        \item $\nu_{\text{bg}}, \nu_{\text{tg}}, \nu_{\text{QPC}}$ are negative and $\nu_{\text{bg}}<\nu_{\text{QPC}}<\nu_{\text{tg}}$, with $V_{\text{tg}}, V_{\text{bg}}<0$,
        
        \item $ \nu_{\text{bg}}, \nu_{\text{QPC}}, \nu_\text{tg}$ are negative and $\nu_{\text{tg}}<\nu_{\text{bg}},\ \nu_{\text{tg}}<\nu_{\text{QPC}}$, with $V_{\text{tg}}, V_{\text{bg}}<0$,

        \item $ \nu_{\text{bg}}, \nu_{\text{QPC}}, \nu_\text{tg}$ are negative and $\nu_{\text{tg}}<\nu_{\text{bg}},\ \nu_{\text{tg}}<\nu_{\text{QPC}}$, with $V_{\text{tg}}<0, V_{\text{bg}}>0$,
        
        \item $ \nu_{\text{bg}}, \nu_{\text{QPC}}$ are positive, $\nu_\text{tg}$ is negative and $\nu_{\text{tg}}<\nu_{\text{QPC}}<\nu_{\text{bg}}$, with $V_{\text{tg}}<0, V_{\text{bg}}>0$.
    \end{enumerate}

Here $|\nu_{\text{tg}}|=2$ is fixed for all regions. 
In the equilibration process, both the charge and heat are exchanged upon tunnelling, giving rise to two internal equilibration lengths. When the geometric length $L$ of the edge is larger than the internal equilibration lengths, a steady state is reached. In that case, the edge states behave like a hydrodynamic mode of a given chirality and are characterized by its topologically governed transport properties, namely the electrical and thermal conductances.  Since we have focused our studies on electrical conductance measurements, we consider phenomena due to full charge equilibration. Thereby we assume $L \gg l^{\text{ch}}_{\text{eq}}$, where $l^{\text{ch}}_{\text{eq}}$ is the charge equilibration length. Furthermore, we note that full charge equilibration washes out any effect of edge reconstruction in electrical conductance. This allows for a simplified picture where we can associate a local electrochemical potential with each mode. 

\begin{figure}[htbp]
    \centering
        \includegraphics[scale=1]{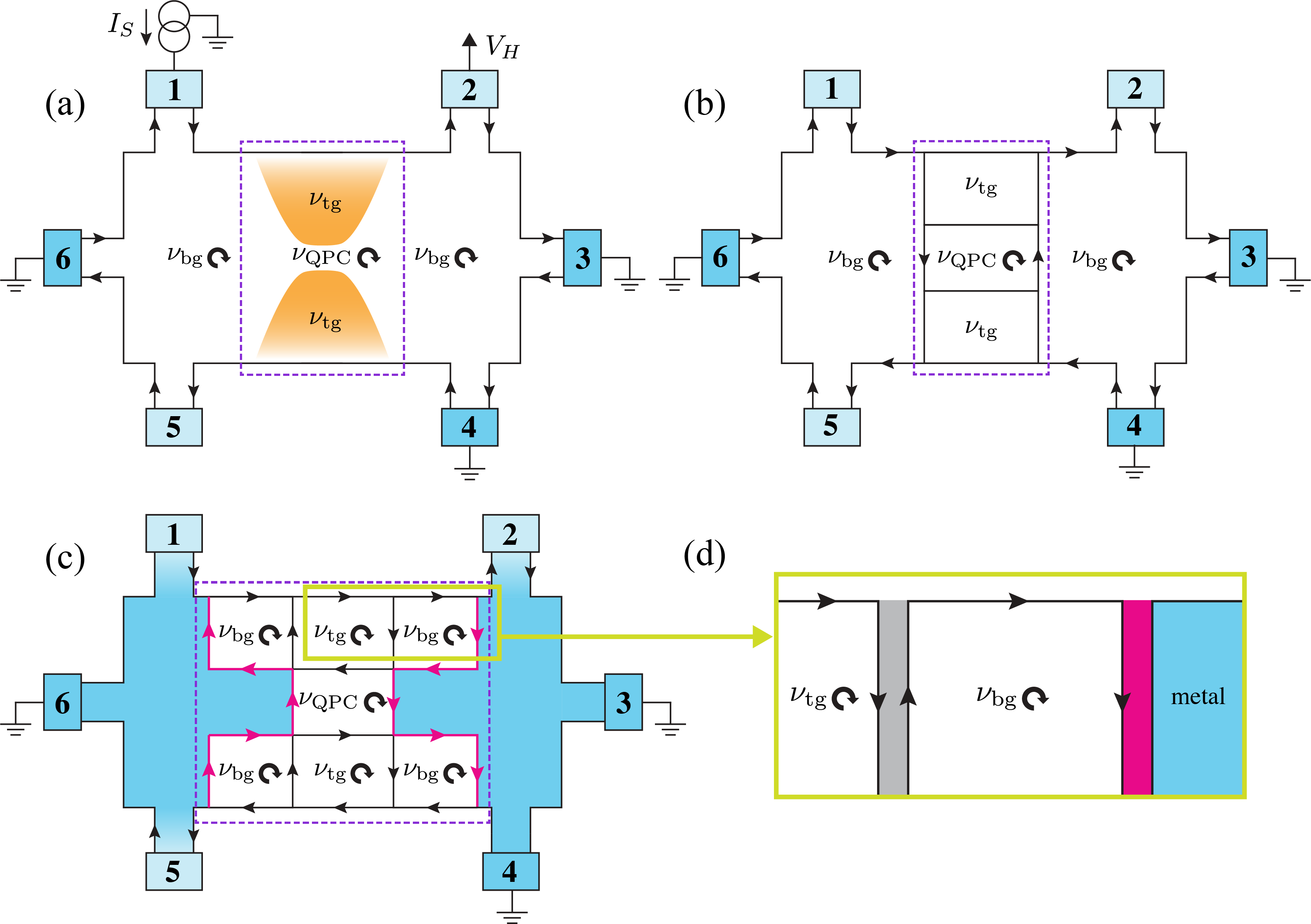}
        \caption{(a) Schematic picture explaining the emergence of the 5/2-plateaus due to the QPC in \text{``Ordinary"} and \text{``Out-of-Ordinary"} scenarios. Metal contacts are colored in blue, where the darker color signifies grounding. Filling fractions with the chirality indicated by the circular arrow are marked in the corresponding area. (b) Edge state propagation for \text{``Ordinary"} regions I (\cref{Fig3}a) and IV (\cref{Fig3}b). The sign of $\nu_{\text{tg}}$ relative to other filling fractions is either the same (region I) or opposite (region IV) and hence the circular arrow for $\nu_{\text{tg}}$ should be adjusted accordingly. (c) Configuration for \text{``Out-of-Ordinary"} regions II (\cref{Fig3}a) and III (\cref{Fig3}b). Here the blue-colored bulk region (gated close to the center of 0th Landau level) behaves like a metal, and edge states are formed only in close surroundings of the top gate. We assume regions close to contacts 1, 2, and 5 exhibit a linear voltage gradient to zero. (d) Close-up of the edges in (c), where the grey region depicts the equilibration between $\nu_{\text{bg}}$ and $\nu_{\text{tg}}$, following the same idea as in \cref{fig:cartoon}. The pink shaded region shows the interface between the metal and $\nu_{\text{bg}}$. It contains many tunneling bridges, allowing the metal and $\nu_\text{bg}$ to exchange particles. Similarly, in the grey region, the two quantum Hall edge states $\nu_{\text{bg}}$ and $\nu_{\text{tg}}$ exchange particles.}
        \label{ManoharFigs}
\end{figure}

\cref{ManoharFigs}a shows a schematic of the QPC Hall bar. The edge state propagation when $\nu_H = 5/2$ is detailed for \text{``Ordinary"} regions I and IV in \cref{ManoharFigs}b and \text{``Out-of-Ordinary"} regions II and III in \cref{ManoharFigs}c. We apply Kirchhoff's current law and assume charge equilibration at each device segment. The Hall resistance is found as $R_H=V_H/I_S$, which for region I leads to 
\begin{equation}
    \nu_H = \frac{|\nu_{\text{bg}}|^2}{|\nu_{\text{QPC}}|},
\end{equation}
and similarly for region IV
\begin{equation}
    \nu_{H}=\frac{|\nu_{\text{bg}}|(|\nu_{\text{bg}}|^2+3|\nu_{\text{bg}}||\nu_{\text{tg}}|-2|\nu_{\text{tg}}||\nu_{\text{QPC}}|)}{2|\nu_{\text{bg}}||\nu_{\text{tg}}|+|\nu_{\text{bg}}||\nu_{\text{QPC}}|-|\nu_{\text{tg}}||\nu_{\text{QPC}}|}.
\end{equation}
Derivations can be found in Supplementary \cref{Note6a}. Using $|\nu_{\text{bg}}|= 2+2/5$ and $ |\nu_{\text{QPC}}|= 2+1/3$, both equations result in $\nu_H \approx 2.47$, close to the experimentally observed $\nu_H =2.5 (\pm 0.05)$.

In \text{``Out-of-Ordinary"} regions II and III, the bottom gate keeps the bulk of the system metallic, seen by the blue color in \cref{ManoharFigs}c, while the portion under the top gate hosts filling $\nu_{\text{tg}}$. Due to the finite potential gradient from the top gate, edge states are stabilized in narrow strips surrounding the split gates, both in the direction of the bulk and inside the QPC \cite{PhysRevB.56.2012,PhysRevB.104.115416}. 
\cref{ManoharFigs}d shows a zoomed-in portion 
of \cref{ManoharFigs}c, where we depict the 
stabilization of a quantum Hall domain at filling
$\nu_{\text{bg}}$ due to the presence of $\nu_{\text{tg}}$. 
We assume $|\nu_{\text{tg}}|>|\nu_{\text{bg}}|$,
following a cascading behavior.
We note that the sign and relative inequalities among the filling factors in region III are effectively the same as in
region II despite the back gate voltage changing sign. This is due to the vanishing of the additional domains arising in region III, which can be explained as follows. Between the top gate and the bulk, the following domains are sequentially formed: an area with a quantum Hall filling that has the same sign as $\nu_{\text{tg}}$, a metallic area, and another quantum Hall portion with
the opposite sign. The latter lies between two metallic domains and is therefore not stable, whereas the first stabilizes, which we denote as $\nu_{\text{bg}}$. Then the picture is equivalent to that of region II where $\nu_{\text{tg}}$ and $\nu_{\text{bg}}$ share the same sign.

The interface between these 
two fillings is equilibrated following the same 
tenet as discussed in \cref{fig:cartoon}.
Similar arguments hold for the emergence of $\nu_{\text{QPC}}$. The resulting charge propagation leads to
\begin{equation}
    \nu_{H}=\frac{|\nu_{\text{bg}}|(|\nu_{\text{tg}}|-|\nu_{\text{QPC}}|)(|\nu_{\text{tg}}|(|\nu_{\text{bg}}|+3|\nu_{\text{tg}}|)-(|\nu_{\text{bg}}|+|\nu_{\text{tg}}|)|\nu_{\text{QPC}}|)}{2|\nu_{\text{tg}}|^3}.
\end{equation}
For $|\nu_{\text{bg}}| = 1+2/3$ and $|\nu_{\text{QPC}}| = 1/3$, we find $\nu_{H}\approx 2.45$. The expected QPC and back gate filling fractions for $|\nu_H|=5/2$ in all four regions are given in \cref{tab:observed_explained}. In Supplementary \cref{Table2}, parent filling fraction pairs $\{\nu_{\text{bg}}, \nu_{\text{tg}}\}$ for other observed $\nu_H$ are marked.
Notably, our calculations are also valid
while changing the overall chirality of each filling and 
thereby adjusting the corresponding inequalities.
\begin{table}[htbp]
\centering
\begin{tabular}{||c| c| c| c||} 
 \hline
 $|\nu_{H}|$ & $|\nu_{\text{tg}}|$ & $\{|\nu_{\text{bg}}|, |\nu_{\text{QPC}}|\}|$  & $\text{Region}$ \\ [0.5ex] 
 \hline\hline
 2+1/2 & 2 & \{2+2/5, 2+1/3\} & I (``Ordinary") \\ 
  \hline
 2+1/2 & 2 & \{1+2/3, 1/3\} & II (``Out-of-Ordinary") \\ 
 \hline
  2+1/2 & 2 & \{1+2/3, 1/3\} & III (``Out-of-Ordinary") \\ 
 \hline
  2+1/2 & 2 & \{2+2/5, 2+1/3\} & IV (``Ordinary") \\[1ex] 
 \hline
\end{tabular}
\caption{The observed quantized conductances $|\nu_H| = 5/2$ shown in \cref{Fig3} with the expected back gate and QPC filling fractions $\{|\nu_{\text{bg}}|$, $|\nu_{\text{QPC}}|\}$.}
\label{tab:observed_explained}
\end{table}

\section*{Discussion}
We present a detailed investigation into the stability of the FQH state with $|\nu_H| = 5/2$ under a variety of external perturbation parameters, exploring two distinct scenarios: the 
``Ordinary" case and the ``Out-of-Ordinary" case.

\textbf{``Ordinary" Case:} \cref{Fig3}c and Supplementary \cref{sevenby2evolve}b illustrate our analysis of $|\nu_H| = 5/2$ in response to variations in magnetic field strength. In both regions I and IV, we observe that reducing the magnetic field induces a transition from $|\nu_H| = 5/2$  to $|\nu_H| = 2 + \frac{2}{3}$. To sustain the stability of the $|\nu_H| = 5/2$ state, a specific combination of filling fractions is required, namely $\{|\nu_{\text{bg}}|, |\nu_\text{QPC}|\}$ = $\{2 + \frac{2}{5}$, $2 + \frac{1}{3}\}$. The systematic evolution of $\nu_\text{QPC}$ from $2 + \frac{1}{3}$ to $2 + \frac{2}{3}$ is linked to the stabilization of a sequence of fractional states as the filling fraction diminishes with increasing magnetic field strengths. Moreover, modulating the top gate filling from $\nu_{\text{tg}} = -3$ to $\nu_{\text{tg}} = -2$ further facilitates the stabilization of $\nu_H = -5/2$, as shown in Supplementary \cref{Gatedependence_1}a-e, thereby corroborating our theoretical predictions based on a full charge equilibration model. We also anticipate that elevating the temperature will enhance equilibration processes, causing the filling fraction in the narrow QPC region, $\nu_\text{QPC}$, to converge towards the back gate filling fraction $\nu_{\text{bg}}$. This phenomenon is evident in Supplementary \cref{FigTevolve_nu5by2}, where $\nu_H = 5/2$ evolves into $\nu_H = 3$ with increasing temperature.

\textbf{``Out-of-Ordinary" Case:} In contrast to the ``Ordinary" case, the state $|\nu_H| = 5/2$ in the ``Out-of-Ordinary" case, emerges via a resonant behavior as the magnetic field strength decreases (see Supplementary \cref{sevenby2evolve}b). This behaviour can be elucidated through the magnetic length $l_B$; as the magnetic field increases, edge states beneath the top split gates become more localized, leading to random tunnelling between the metallic bulk and the edge states of the top split gate, thus yielding a finite-size feature. When the magnetic field is reduced, the edge states under the split top gate extend more toward the bulk states, facilitating improved equilibration and producing a more robust plateau at $|\nu_H| = 5/2$. Similar to findings in the \text{``Ordinary"} case, adjusting the filling fraction of the top split gate to $\nu_{\text{tg}} = -2$  stabilizes $\nu_H = -5/2$, as depicted in Supplementary \cref{Gatedependence_2}. However, the temperature dependence in the  \text{``Out-of-Ordinary"} case contrasts sharply with that observed in the ``Ordinary" case. At lower temperatures, $\nu_H = 2 + \frac{1}{3}$ is observed (see \cref{Fig3}d), and with the increase of the electronic temperature the transition to $\nu_H = 5/2$ is seen. This transition is likely due to increased metallicity within the QPC region, facilitated by enhanced equilibration dynamics with the metallic region in the bulk filling at higher temperatures. Our comprehensive equilibration model effectively accounts for the observed FQH states and their response to various external tuning parameters.

\section*{Summary and Outlook}
In this work, we observed half-quantized fractional quantum Hall plateaus in the confined geometry of monolayer graphene. These plateaus appear for more than one combination of the top and back gate electrostatic potentials. The resulting mechanism leading to the curious observation of the conductance plateau at $\nu_H = 5/2$  can be classified into two cases: i) ``Ordinary" tunneling between the FQH fluid with different filling factors, and ii) ``Out-of-Ordinary" FQH fluid in contact with Fermi-liquid-like reservoirs. For both cases, charge equilibration between parent states leads to the half-quantized plateaus. In that regard, these can be considered as ``designer" states with tunable charge. This raises an important question about the role of equilibration in quantum Hall measurements, specifically in confined geometries. 

Here, it is necessary to note that while charge equilibration allows for these exotic states, it also destroys the coherence of the system. Mitigating this, thermal quantum Hall transport can be considered as an alternative for probing designer states since the thermal equilibration length is one order of magnitude larger than the charge equilibration length. Nevertheless, the exact nature of the unconventional 5/2 plateau is yet to be understood. As a notable consequence of edge equilibration, voltage drops occur due to carriers encountering metallic contacts or other charge carriers with different electrochemical potentials, giving rise to hot spots in the sample. Power is dissipated at these locations due to Joule heating, which causes heat to propagate along the edge modes, creating particle-hole pairs. These thermally activated pairs start to tunnel across edge modes, and a stochastic partition of these, reaching different contacts, leads to shot noise. Measuring this would give a more detailed picture of the non-equilibrium processes in the QPC region. Moreover, both thermal transport and shot noise can be employed for studying the possibility of edge reconstruction at the plateaus found in our work.

Our current study directs us toward a more careful evaluation and understanding of confined geometry. Being able to tune the charge of states due to the different QPC filling opens up possibilities for anyon collider experiments utilizing QPCs to generate diluted beams. The cross-correlation between two such beams can probe the details of the quantum statistics of these particles, bringing us one step closer to building topological qubits using fractional quantum Hall states.

\section*{Methods}
\subsection*{Sample fabrication}
A standard dry transfer technique was used to fabricate hexagonal boron nitride (hBN) encapsulated graphene samples with graphite back gates. Both hBN and graphene crystals were exfoliated on 280 nm thick SiO$_2$/Si(p++) substrates. Suitable hBN and graphene crystals were first identified through optical microscopy, and additional RAMAN spectroscopy was performed to select monolayer graphene. Selected crystals were then successively picked using a polycarbonate/polydimethylsiloxane stamp to create the desired hBN/graphene/hBN/graphite heterostructure, and later dropped on a clean sapphire substrate. The clean and bubble-free region of the fabricated stack was identified using atomic force microscopy. The edge contacts to monolayer graphene were fabricated in a self-aligned manner where polymethyl methacrylate was used as a standard e-beam resist mask, followed by CHF$_3$ + O$_2$ reactive ion etching to etch the stack and expose the 1D edges of graphene \cite{Wang2013}. The same resist mask was then used to metalize the exposed graphene edges and form the 1D contacts made of Ti/Au, where a 5 nm Ti layer is used as an adhesive layer between the graphene and Au contact. The top split gates defining the QPC were fabricated in a second lithography step by patterning over the hBN/graphene/hBN/graphite stack, followed by Ti/Au metal deposition. The lithography of the top split gate was done by $100$ keV e-beam, which may cause lattice defects under the gate, resulting in strong equilibration in the confined region of QPC.   The top split gate has a spacing of $60-80$ nm. Contacts to the bottom graphite gate were also defined in the same step. In the final step, Hall bar structures were cut out with e-beam lithography followed by a second etch with CHF$_3$ + O$_2$. The final structure of a device with a typical size of 6 \textmu m $\times$ 2.5 \textmu m is shown in \cref{Fig1}a.   
\subsection*{Measurement techniques}
The magneto-transport measurements are performed in a Bluefors bottom-loading fast-sample exchange system with a base temperature of $18$ mK. The sample exchange platform is equipped with two second-order low-pass filters. Separate stainless steel powder filters are installed at the mixing chamber to reject microwave frequencies. The effective cut-off frequency for all measurement lines is $10$ kHz. This effective filtering ensures the electronic temperature remains close to the base temperature of the cryostat. The transport measurements were performed using the standard lock-in techniques with a frequency of $13.333$ Hz and a current of $1$ nA. Two Zurich MFLI amplifiers were used in sync mode to measure the Hall and longitudinal voltages. Four samples were fabricated and characterized for this experiment. The gate response of the Hall bar device is shown in Supplementary \cref{fig:comsol}c. All four samples show the charge neutrality point close to zero back gate voltage with slightly negative doping $V_{bg} \simeq -0.05$  to $-0.1$ V. For the top gate, the peak in resistance, \textit{i.e.} the Dirac point is seen at $V_{tg} = -0.75$ to $-0.1$ V. The resistances $R_{xy} = V_{2-4}/I_S$ and $R_{xx} = V_{1-2}/I_S$ are measured with the bias current $I_S$ of $1$ nA fed at contact 6. Here contact $3$ is a cold ground. $R_H = V_{2-3}/I_S$ and $R_\text{ref}=V_{5-6}/I_S$ are measured with $I_S$ applied to contact 1, while keeping contacts 3, 4 and 6 grounded. We note that $R_{xy}$ and $R_H$ are equivalent due to the grounding at contact 3. The measured filling factor $\nu_H$ is therefore defined by $R_H=h/(\nu_He^2)=R_{xy}$. The measurements were focused in high magnetic field range: $B = 10-13.5$ T. All data presented in this report, including the supplementary materials, are from Hall bar samples SH4: \cref{Fig1}c, \ref{Fig3}a-c, Supplementary \cref{fig:comsol} and \ref{HallR}, and SH5: \cref{Fig1}d, \ref{Fig2}a-c, \ref{Fig3}d, Supplementary \cref{fig:note4}, \ref{FigTevolve_nu5by2}, \ref{Gatedependence_1}, \ref{Gatedependence_2}, \ref{sevenby2evolve}, \ref{gapenergy_1} and \ref{gapenergy_2}.     

\section*{Acknowledgement}
 We acknowledge funding from the Research Council of Finland projects 312295 and 352926 (CoE, Quantum Technology Finland) as well as grant 338872 (NAI-CoG). Our work was also supported by the Ministry of Education of Finland via the Finnish Indian Consortia for Research and Education (FICORE). Our work is part of the QuantERA II Program that has received funding from the European Union’s Horizon 2020 Research and Innovation Programme under Grant Agreement Nos 731473 and 101017733. We are grateful for fruitful discussions with Sudhansu Mandal, Ajit Balram, Arup Kumar Paul, and Jakub Tworzyd\l{}o. We thank the International Centre for Theoretical Sciences (ICTS) for participating in the program - Condensed Matter meets Quantum Information (code: ICTS/COMQUI2023/9), from where the collaboration originated. S.M.\ was supported by the Weizmann Institute of Science, Israel Deans fellowship through Feinberg Graduate School. A.D. thanks the IISER Tirupati start-up grant for support. Our experimental work benefited from the Aalto University OtaNano/LTL and OtaNano/NanoFab infrastructures.

\section*{Author contributions}
P.P. designed the sample with hBN provided by K.W. and T.T. P.P. performed the experiments including the data collection. M.K., P.P., and K.N.F. performed the data analysis. K.N.F. developed the electrostatic model for the QPC Hall bar. A.D. and A.S. developed the bottom loading puck for the cryostat. A.D. developed the theoretical model and S.M. performed the theoretical calculations. The results and their interpretation were discussed with P.P., S.M., A.D., P.J.H., K.N.F., and J.S. All authors participated in writing the article. M.K. supervised the project.    
\section*{Competing interests} The authors declare that they have no competing interests.
\section*{Data availability}
All data needed to evaluate the conclusions in the paper are present in the report and the supplementary materials. Additional data related to this paper may be requested from the corresponding author.

\newpage
\setcounter{figure}{0}
\setcounter{table}{0}
\renewcommand{\thefigure}{S\arabic{figure}}
\renewcommand{\thetable}{S\arabic{table}}
\section* {Supplementary Information for ``Half-quantized Hall Plateaus in the Confined Geometry of Graphene"}   

\section{Simulation for the effective field in the presence of QPC } \label{Note3}
\begin{figure}[htbp]
    \centering
    \includegraphics[width=0.95\textwidth]{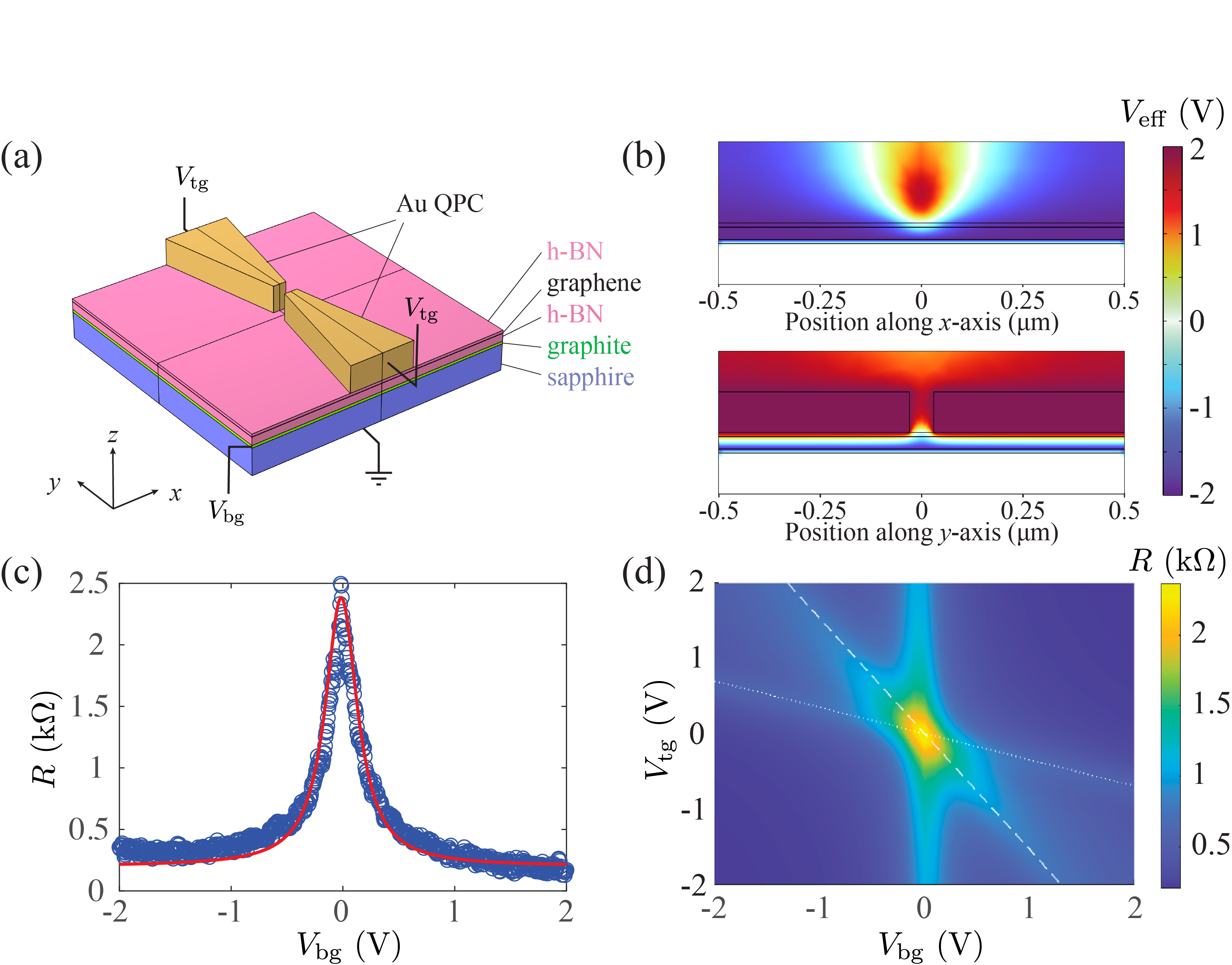}
    \caption{Finite element simulation of the sample. The model is shown in (a). The sample is placed in a box of air, omitted in the figure for clarity. The layer thicknesses used are $d_\text{sapphire}=100$ nm, $d_\text{graphite}=10$ nm, $d_\text{hBN}^\text{bottom} = 30$ nm, $d_\text{graphene}=1$ nm, $d_\text{hBN}^\text{top} = 10$ nm and $d_\text{Au} = 100$ nm. The QPC gap is $d_\text{gap} = 60$ nm. (b) Cross-sectional view of the potential distribution $V_\text{eff}$ along the $x$-axis, centered inside the gap of the QPC (top), and the $y$-axis, which runs through the QPC (bottom). The applied voltage is $V_\text{tg}$ = 2 V and $V_\text{bg}$ = -2 V. $V_\text{eff}$ along both axes at the center of the graphene is mapped to $R_{xx}$ using the fit (red) from the experimental data (blue) for the resistance measurement at $B$ = 0 T and $V_\text{tg}$ = 0 in (c). (d) The values are superimposed for a sweep of $V_\text{tg}$ and $V_\text{bg}$ to $\pm$2 V. The dotted line has slope $m_1=-0.35$ and the dashed line has slope $m_2 = -1.54$.}
    \label{fig:comsol}
\end{figure}

Electrostatic simulations using COMSOL Multiphysics\textsuperscript \textregistered are performed to study the effects of the gates on the graphene. The Laplace equation $\nabla^2V = 0$, where $V$ is the scalar potential, is solved using the finite element method. The device stack is modelled in three dimensions for an area of 1 \textmu m$^2$ centered at the QPC, as seen in \cref{fig:comsol}a. The different material layers are included by appropriate thickness and dielectric constant, where the graphene is represented by a 1 nm thin sheet with a relative permittivity $\epsilon =  4$ \cite{Nikolai2015, siegel2011many}. The top gate $V_\text{tg}$ is applied to the QPC, and the bottom gate $V_\text{bg}$ is applied to the graphite layer. The sapphire substrate is kept at 0 V as the ground terminal. \cref{fig:comsol}b shows the cross-sectional view of the electric potential $V_\text{eff}$ along the $x$-axis, centered inside the gap of the QPC, when $V_\text{tg}$ = 2 V and $V_\text{bg}$ = -2 V. As can be seen in the bottom plot, which shows the distribution along the $y$-axis at $x = 0$, cutting laterally through the QPC, the top gate has a strong influence in its close vicinity. But away from the QPC, shown in the upper plot, its effect dies off and the back gate contributes to the potential for the larger part of the sample.
\par This becomes more apparent when plotting the resistance $R$ as a function of $V_\text{tg}$ and $V_\text{bg}$. A response function $R(V)$ is found from the experimental data (\cref{fig:comsol}c) in terms of the applied voltage $V$, which is described by a Lorentzian curve of the form
\begin{equation}
    R(V) = R_0+\frac{2A}{\pi}\frac{w}{4(V-V_D)^2+w^2},
\end{equation}
where $R_0$ is a baseline offset, $A$ is the area under the curve, $w$ is the width at half maximum and $V_D = -0.015$ V is the center of the peak. $R(V_\text{eff})$ is calculated for $n = 1001$ points along the $x$ and $y$-axis inside the graphene. Since the QPC has a strong effect on the longitudinal transport, the points on the $y$-axis are averaged such that 
\begin{equation}
    R\big(V_\text{eff}^{(y)}\big) = \frac{\sum_{i=1}^n \alpha(y_i) R(V_\text{eff}^{(y_i)})}{\sum_{i=1}^n \alpha(y_i)}.
    \label{eq:average}
\end{equation}
Here $\alpha(y_i)$ is calculated using a transmission probability set to $\gamma$ = 0.2 below the top gates and $1-\gamma$ = 0.8 in the gap. The value in the gap is scaled by a factor of 10 to account for the weighting based on the number of points. This gives
\begin{equation}
   \alpha(y_i) = -\frac{10(1-\gamma)-\gamma}{1+e^{-6(100|y_i|-3)}}+10(1-\gamma).
\end{equation}

\par The final resistance map is found as $R(V_\text{bg}, V_\text{tg})= \frac{1}{2}\big[R\big(V_\text{eff}^{(x)}\big) + R\big(V_\text{eff}^{(y)}\big) \big]$, shown in \cref{fig:comsol}d. It shows the dominant effect from the bottom gate with the vertical line at $V_\text{bg} \simeq 0$. A weaker contribution from $V_\text{tg}$ can be seen in the diagonal lines, where two distinct slopes can be identified. These correspond to $m_1  = -0.35$ and $m_2 = -1.54$, which are similar to the slopes $m_\text{tg} = \Delta V_{tg}/\Delta V_{bg} = -0.36$ and $m_\text{QPC} =\Delta V_{tg}/\Delta V_{bg} = -1.55$ found from the data (see \cref{Fig1} in the main text and \cref{fig:note4}).

\section{Quantum Hall transport measurement} \label{Note6}
\begin{figure}[htbp]
    \centering
    \includegraphics[]{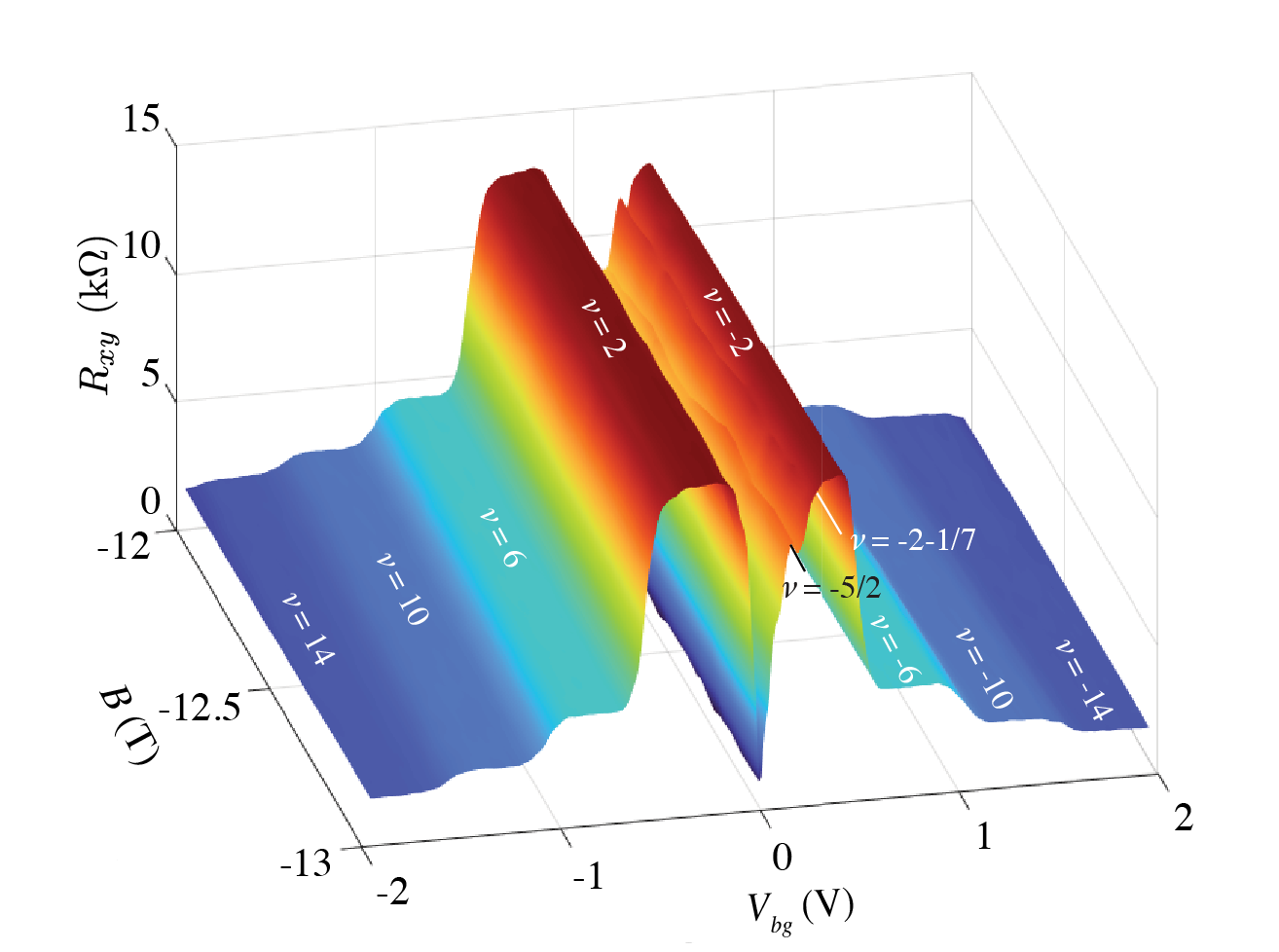}
                  \caption{$R_{xy}$ characterization of Hall bar sample at high magnetic fields. The quantum Hall plateaus dominantly seen here are at filling factors $|\nu_H|= 2, 6, 10$ and 14. Higher filling factors ($|\nu_H| = 18, 22, 26, 30, 34$, and $38$) are seen at lower magnetic fields (not shown here).  Remarkably, fractional quantization is seen at $\nu_H \simeq -2-1/2$ and $-2-1/7$ when the back gate is tuned from Dirac voltage to higher positive gate voltage. Note that the QPC is kept at the charge neutrality point $V_\text{tg} = -0.75$ V for this measurement.
                  }
    \label{HallR}
\end{figure}

The surface plot of $R_{xy}$, matching $R_{xx}$ shown in \cref{Fig1} in the main text, can be seen in \cref{HallR}. Hall plateaus with filling factors $|\nu_H|= 2, 6, 10$ and 14 are observed. Additionally, steps at $\nu_H \simeq$ -5/2 and -2-1/7 emerge when $B \sim 13$ T. With the reduction of the magnetic field, these plateaus evolve to -2-2/3 and -2-1/5, respectively.

\section{Analysis of filling fraction in QPC region} \label{Note4}
\begin{figure}[htbp]
    \centering
    \includegraphics[scale=1]{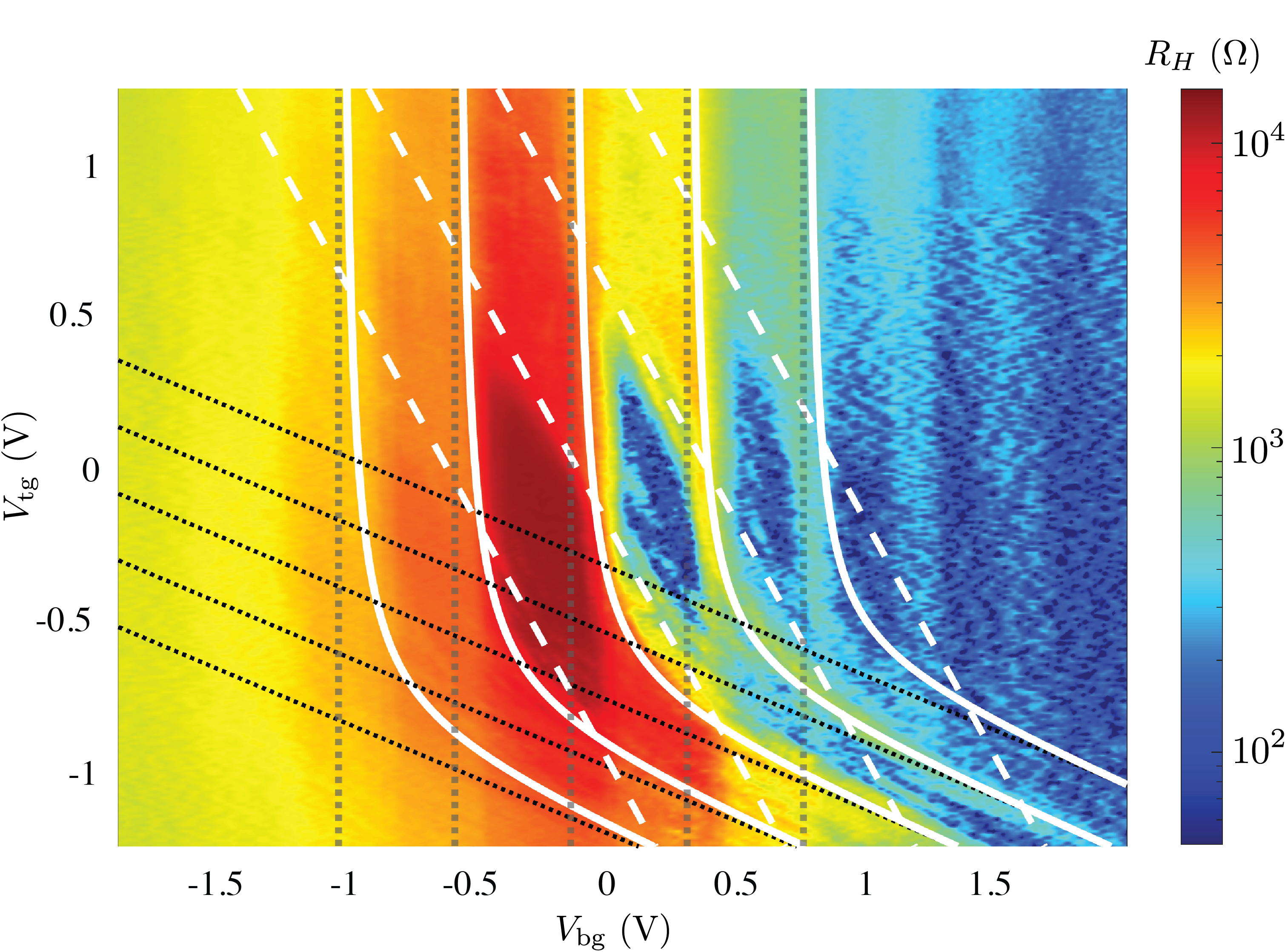}
    \caption{Colormap of $R_{H}$ at $B =$ 10 T. White dashed lines display slope $m_\text{QPC} =-1.55$. The black dotted lines correspond to slope $m_\text{tg}=-0.36$. White solid lines show $f^{(j)}(x)$ with asymptotes indicated by the grey and black dotted lines.}
    \label{fig:note4}
\end{figure}

\cref{fig:note4} shows a colormap of $R_{H}$ at $B$ = 10 T. The resistance forms rhombi patterned by vertical (grey dotted) and slant lines (white dashed) with slope $m_\text{QPC}=-1.55$. Towards the lower part of the plot, another slope $m_\text{tg}=-0.36$ (black dotted lines) becomes visible. Minima in the resistance can be found along curves shaped by asymptotes corresponding to the black and grey dotted lines. This shows the modulation of the charge density from the combination of the back and top gates. Five curves (white solid lines) are shown in the plot. Let $f(x)$ represent $V_\text{tg}$ with $x$ given by $V_\text{bg}$. Then the formula for the $j = 1,..,5$ curves is given by

\begin{equation}
    f^{(j)}(x) = a\frac{\big(x-x_1^{(j)}\big)\big(x-x_2^{(j)}\big)}{x-v^{(j)}},
\end{equation}
where $a$ is a stretch factor, $x_1^{(j)}$ and $x_2^{(j)}$ are the $x$-intercepts, one of which is of a mirrored curve (not shown), and $v^{(j)}$ is the position of the vertical asymptote. The positions can be found as 
\begin{align}
    \begin{split}
        &v^{(j)} = -1.5+0.45j, \\
        &x_1^{(j)} = v^{(j)} + d_1 \text{ with } d_1 = 0.08, \\
        &x_2^{(j)} = -\frac{c^{(j)}}{m}-d_2, \\
        &c^{(j)} = -1.43+0.22j, \\
        &d_2 = \frac{d_1^2}{\sin\big(|\tan^{-1}(m)|\big)}.
    \end{split}
\end{align}
The stretch factor is constant for each curve, found at a far-away point $x = 100$ by applying $y = mx+c^{(j)}$ such that 
\begin{equation}
    a = \frac{y\big(x-v^{(j)}\big)}{\big(x-x_1^{(j)}\big)\big(x-x_2^{(j)}\big)}  = -0.3596.
\end{equation}

\section{Operation of QPC in graphene in quantum Hall regime} \label{Note4a}
The top split-gate geometry is the foundational tool for engineering quantum point contacts (QPC) in the bandgap semiconductor GaAs/AlGaAs. However, the QPC in graphene differs from that in GaAs/AlGaAs. Graphene is semimetal; thus, by applying a potential on the top gates, the carriers underneath can be modulated from electron to hole type or vice versa. Hence, with the combination of the back gate and top gate, the graphene regions form either a unipolar regime ($nn'n$ or $pp'p$) or a bipolar regime ($pn'p$ or $np'n$). The $pn$-junction interface defined by the top gate allows for unconventional tunneling of particles via Klein tunneling processes at zero magnetic fields.  

\par On the other hand, at high magnetic fields, Landau levels are formed and electron transport across the QPC region is governed by the chirality and number of edge states in the bulk and under the top gate. In the unipolar regime, There are two cases: (i) $\nu_{\text{bg}} < \nu_{\text{tg}}$ and (ii) $\nu_{\text{tg}} < \nu_{\text{bg}}$. 
\par
\begin{figure}[bhtp]
    \centering
    \includegraphics[scale=1]{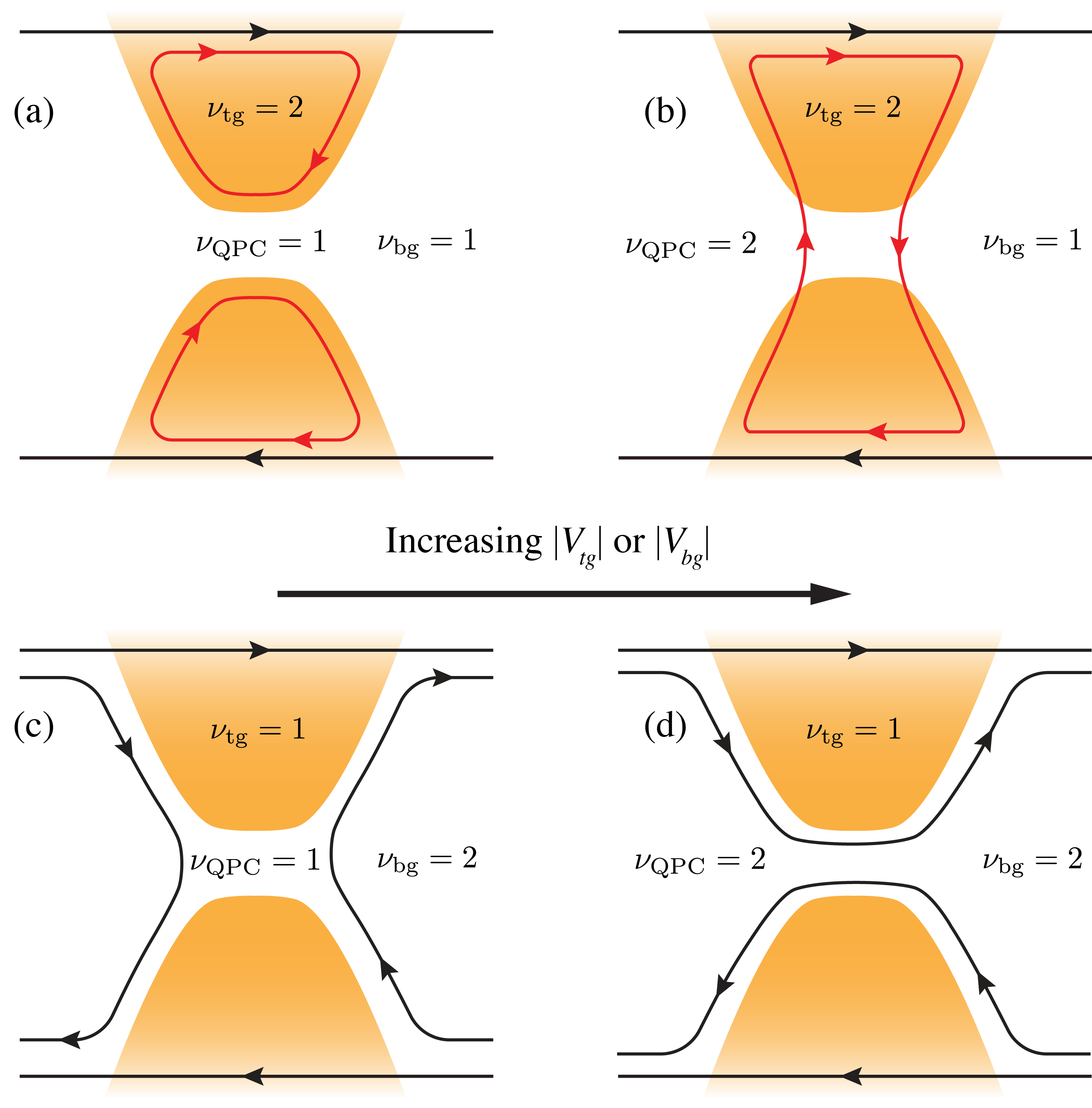}
    \caption{Schematic of edge states in QPC geometry in unipolar regime: (a) and (b) $\nu_{\text{bg}} \leq \nu_{\text{QPC}} \leq \nu_{\text{tg}}$, (c) and (d) $\nu_{\text{tg}} \leq \nu_{\text{QPC}} \leq \nu_{\text{bg}}$. The slight increase of potential on either the top gate or bottom gate leads to an increase in the filling factor in the QPC region. The edge states $\nu = 1$ and $\nu = 2$ have different spin polarization; thus, there will be no equilibration processes in either of the four cases. The resultant $R_{H} = h/e^2$ for case (a)-(b) and $R_{H} = h/2e^2$ for (c)-(d).}
    \label{fig:note6}
\end{figure}

 For case (i) when $\nu_{\text{bg}} < \nu_{\text{tg}}$, the excess edge states ($\nu_{\text{tg}} - \nu_{\text{bg}}$) form the localized states under the top gate and edge channel $\nu_{\text{bg}}$ defined by the back gate passed through the top gate region unhindered (see \cref{fig:note6}a). With tuning of the top gates, \textit{i.e.} increasing $\nu_{\text{tg}}$, the localized states may extend over the QPC region (see \cref{fig:note6}b). Nevertheless, this will not affect the transmitted $\nu_{\text{bg}}$ edge states. Hence, without any equilibration processes, $R_H = h/(\nu_{\text{bg}} e^2)$, as shown in \cref{fig:note6}. 

\par For case (ii) when $\nu_{\text{tg}} < \nu_{\text{bg}}$, the edge channel with filling $\nu_{\text{bg}} -\nu_{\text{tg}}$ follows the equipotential line marked by the top gate and passes through the QPC regime (see \cref{fig:note6}c and d). Here the edge channel defined by $\nu_{\text{tg}}$ passes unhindered through the top gated region. Again, without interaction between the two counter-propagating edge states, $R_H = h/(\nu_{\text{bg}}e^2$). 

\par Similarly, the transmission across the QPC, \textit{i.e.} the diagonal resistance $R_D$ is defined by the number of channels crossing through under the top gates and edge states circulating across the split top gates. In both cases (i) and (ii), ideally $R_D = e^2/h\nu_{\text{QPC}}$. Despite that, in the presence of the equilibration, both $R_H$ and $R_D$ may deviate from their ideal values, the details of which lie in the edge dynamics of the QPC. Thus an additional parameter, the local filling factor in the QPC region $\nu_{\text{QPC}}$ defines the dynamics of the edge states. For case (i), the following condition should be followed: $\nu_{\text{bg}} \leq \nu_{\text{QPC}} \leq \nu_{\text{tg}}$. The number of circulating edge states between the two top gates is then $\nu_{\text{QPC}} - \nu_{\text{bg}}$ and the number of localized states under the top gates is $\nu_{\text{tg}} - \nu_{\text{QPC}}$. For case (ii), condition $\nu_{\text{tg}} \leq \nu_{\text{QPC}} \leq \nu_{\text{bg}}$ should be satisfied. The number of edge states reflected from the QPC is then given by $\nu_{\text{bg}} - \nu_{\text{QPC}}$. 

\par The backscattering between the two counter-propagating channels could happen due to equilibration processes between the transmitting edge channels with extended localized states over the QPC region or back-reflected edge channels. This equilibration process relaxes edge states to a new equilibrium, eventually redistributing the current among the edge channels with the same chemical potential. 

\par For example, if spin polarization is lifted first in N = 0 and N =1 Landau level, the edge states with filling factors $\nu = 1$ and $\nu = 3$ and $4$ will have the same spin polarization, and hence they may equilibrate, despite being $\nu = 3$ and $\nu = 4$ belonging to different valleys. On the other hand, if valley polarization is lifted first, despite both $\nu = 4$ and $3$ having the same valley polarization, they will not equilibrate owning to their different spin polarization. Hence, contrary to the previous case, only $\nu = 3$ will equilibrate with $\nu =1$. Thus the detailed outcomes of equilibration depend on the order in which spin and valley degeneracies are lifted.

\par So far in experiments, only spin-selective equilibration processes have been observed. The absence of valley polarization in the equilibration process may be related to valley mixing due to inter-valley scattering from lattice defects. Still, even in high-mobility samples, the conservation of valley polarization has not been observed in the equilibration processes. Thus, it is still an open question.

\par
\begin{figure}[bhtp]
    \centering
    \includegraphics[scale=0.6]{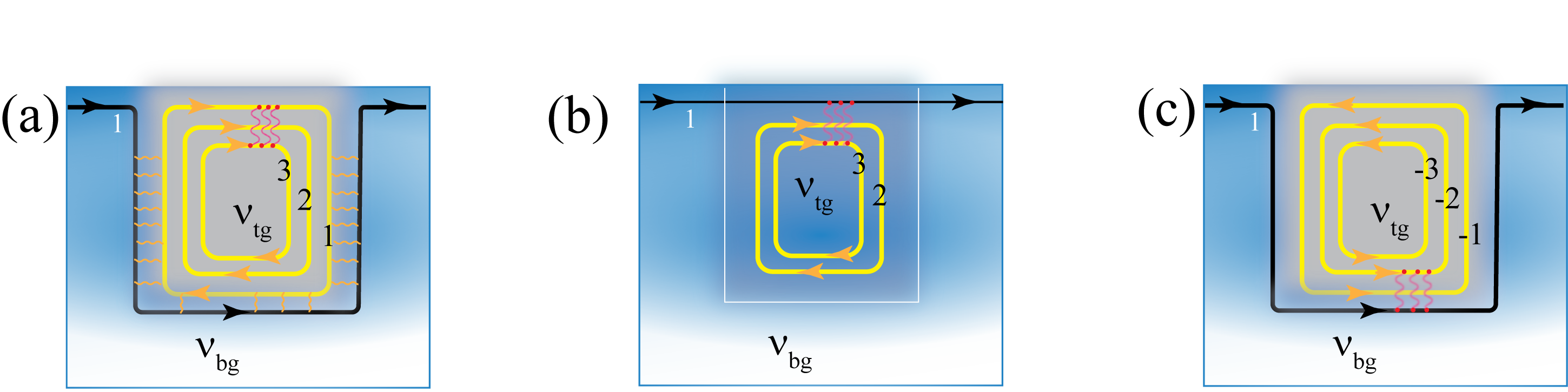}
    \caption{Equilibration processes in quantum Hall edge channels in graphene Hall bar with QPC. Edge states undergo selective spin equilibration if the spin and valley symmetry is broken: (a) and (b) Unipolar regime: $\nu_{\text{tg}} > \nu_{\text{bg}}$. Here $\nu = 1$ under the top gate (yellow) and $\nu = 1$ under the back gate (black) form counter-propagating states in their respective region. These two counter-propagating states will be gapped to form the continuous edge (black line). Also, the spin polarization of $\nu = 1$ and $\nu = 3$ are the same and thus undergo equilibration. (c) Bipolar regime: $|\nu_{\text{tg}}| > |\nu_{\text{bg}}|$ and $\nu_{\text{tg}} < 0$.  Also, the edge states in the top gate region and the bulk area defined by the back gate will form co-propagating edge states, so they will not be gapped. Due to identical spin states of $\nu = 1$ and $\nu = -2$, they will equilibrate.}
    \label{fig:note6b}
\end{figure}

 In the unipolar case of $\nu_{\text{bg}} \leq \nu_{\text{tg}}$, for example when $\nu_{\text{bg}} = 1$ and $\nu_{\text{tg}} = 3$, there will be finite equilibration between the edge channels (see \cref{fig:note6b}a). Since the edge channel with filling factor $\nu_H = 3$ is localized under the top gate, there will be no loss on total transmission across the QPC. On the other hand, if the edge channel ($\nu_H = 3$) is extended over the QPC region, there will be finite interaction between two counter-propagating edge channels ($\nu_H = 1 $) via an extended edge state ($\nu_H = 3$) in QPC region which will lead to a reduction in the transmitted current.  However, if  $\nu_{\text{tg}} = 3$, $\nu_{\text{bg}} = 1$ and $\nu_{\text{QPC}} = 2$, then an edge channel with filling factor $\nu_H = 2$ will extend over the QPC region. In this case, since $\nu_H = 1$ and $\nu_H = 2$ don't have the same spin state, there will be no equilibration and hence no reduction in the QPC transmission. A robust plateau at $R_H = h/e^2$ will be seen for the filling factor in the QPC region 
within the range of $1\leq \nu_{\text{QPC}} \leq 2$. And with the increase of $\nu_{\text{QPC}} \geq 2$, the resistance
will increase further and thus deviate from the standard value $h/e^2$. These processes are not limited to the lowest Landau Levels $N = 0$, they may happen even at higher Landau levels $N > 0$. But, with the increase in the number of Landau levels, the screening effect will exponentially suppress such processes for the innermost edge channels. 

\par
\begin{figure}[htbp]
    \centering
    \includegraphics[scale=1]{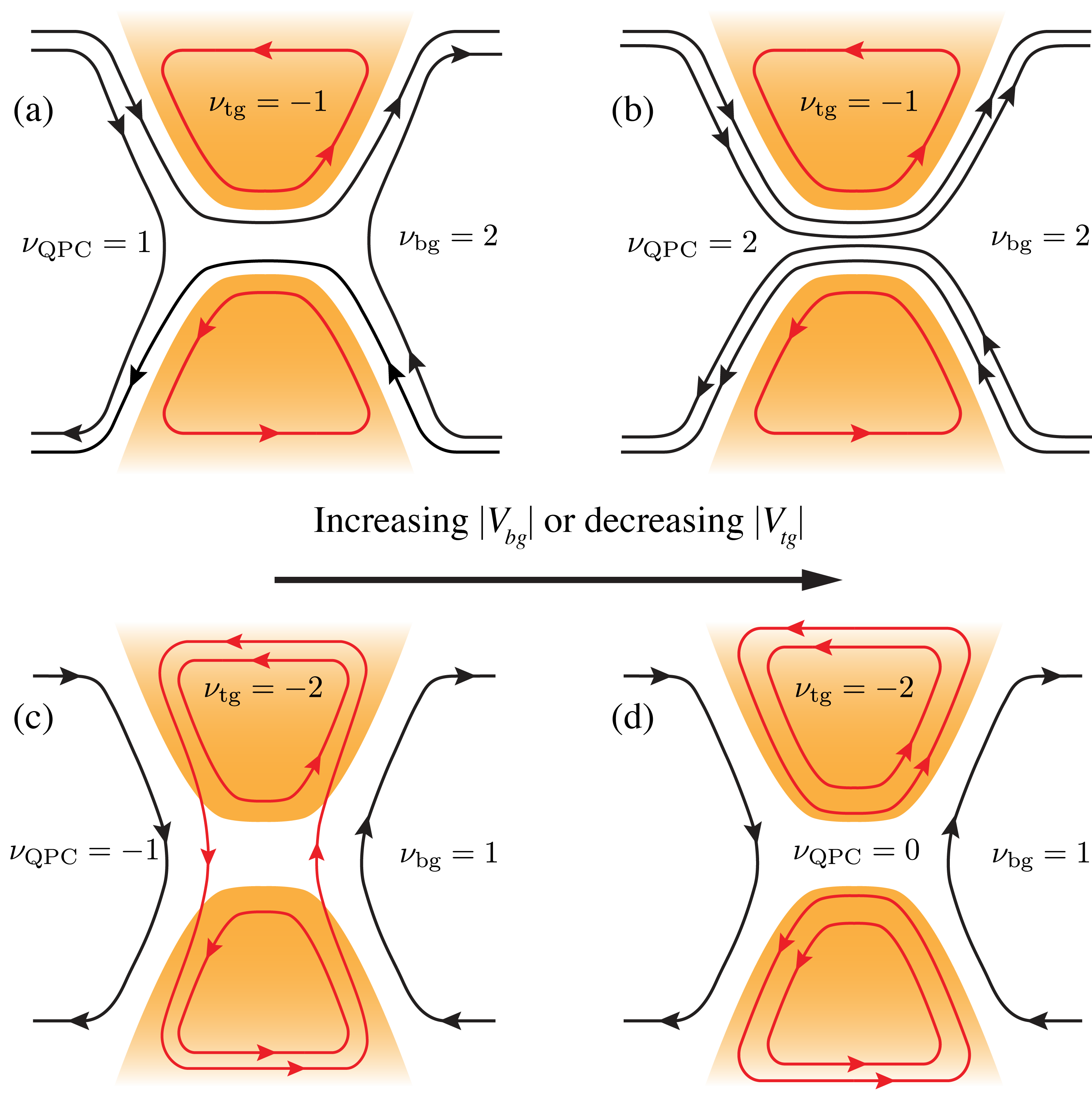}
    \caption{Schematic of edge states in QPC geometry in bipolar regime with $\nu_{\text{tg}} < 0$ and $\nu_{\text{bg}} > 0$: (a) and b) $|\nu_{\text{tg}}| \leq |\nu_{\text{QPC}}| \leq |\nu_{\text{bg}}|$, (c) and (d) $|\nu_{\text{QPC}}| \leq |\nu_{\text{bg}}| < |\nu_{\text{tg}}|$. Changing the potential on either the top gate or bottom gate leads to modulation of the filling factor in the QPC regime. The edge states $\nu = -1$ and $\nu = 2$ have the same spin polarization; there will be equilibration in cases (a) and (b). The effect of equilibration is more evident in (c) and (d) where $\nu_{\text{QPC}} = 0$, indicating a fully pinched QPC. In this case, the only way to transmit the current across the QPC will be equilibration between $\nu = -2$ (under the top gate) and $\nu = 1$ (the backscattered edge channel).}
    \label{fig:note6a}
\end{figure}

 In the bipolar regime, the chirality of the edge states under the top gate and the rest of the graphene due to the global back gate will be different. This results in the formation of the localized states $\sim \nu_{\text{tg}}$ under the top gate. These localized states may extend over the QPC region with the tuning of the top gate voltage.  On the other hand, edge channels defined by the back gate voltage either pass through the QPC regime or with part of them fully reflected (see \cref{fig:note6a}a-d).  

\par Like unipolar regimes, the equilibration only appears for identical spin-polarized states (see \cref{fig:note6b}b). Such processes dominate the lowest Landau level $N = 0$ at higher fields and may happen between $N = 0$ and $N = 1$ states at lower fields.  Note that the QPC could be fully pinched in the bipolar regime only. Thus, for stronger tunability of the QPC, it is more beneficial to operate in the bipolar regime. 

\par For the case of degenerate states, equivalently all edge channels will equilibrate. This is because spin and valley symmetries are not broken; thus, spin and valley quantum numbers need not be conserved. Moreover, screening the innermost edge channel limits the equilibration to only a few outermost edge channels. For example, as shown in \cref{fig:note6b}c, the edge channel $\nu = -2$ equilibrates more likely with $\nu = -6$ than with $\nu = -10$. Similarly,
equilibration also occurs in the FQH regime.

\section{QPC in fractional quantum Hall regime}
Strong tunneling and interaction among the edge modes can occur in the FQH regime, facilitating their equilibration. The smaller difference in the energy gaps leads to faster equilibration than the integer quantum Hall regime.
When multiple quantum Hall modes propagate with their corresponding chiralities, the equilibration mechanism makes them effectively one hydrodynamic mode with a net charge conductance propagating along a given direction (see main text \cref{fig:cartoon}a).
A more involved scenario occurs in our QPC geometry, although the philosophy of charge propagation due to equilibration remains the same, as shown in \cref{fig:cartoon}b. Here we have the case $\nu_{\text{bg}}, \nu_{\text{tg}} \text{ and } \nu_{\text{QPC}}$ are negative and $\nu_{\text{bg}}<\nu_{\text{QPC}}<\nu_{\text{tg}}$; other cases can be considered similarly.
In our sample, the formation of FQH states is evident from the intermediate steps/plateaus observed between two integer quantum Hall plateaus. The edge structure for the FQH states is far more complex than that of their parent integer quantum Hall edge states. However, the equilibration picture makes it simple enough to explain the experimentally found non-trivial $R_{H}$ without loss of generality.

\section{Stability of the plateau at the half filling \texorpdfstring{$\nu_H = 5/2$}{TEXT}} \label{NoteStability}
\subsection{Temperature dependence} \label{Note5a}

\begin{figure}[htbp]
    \centering
    \includegraphics[width=\textwidth]{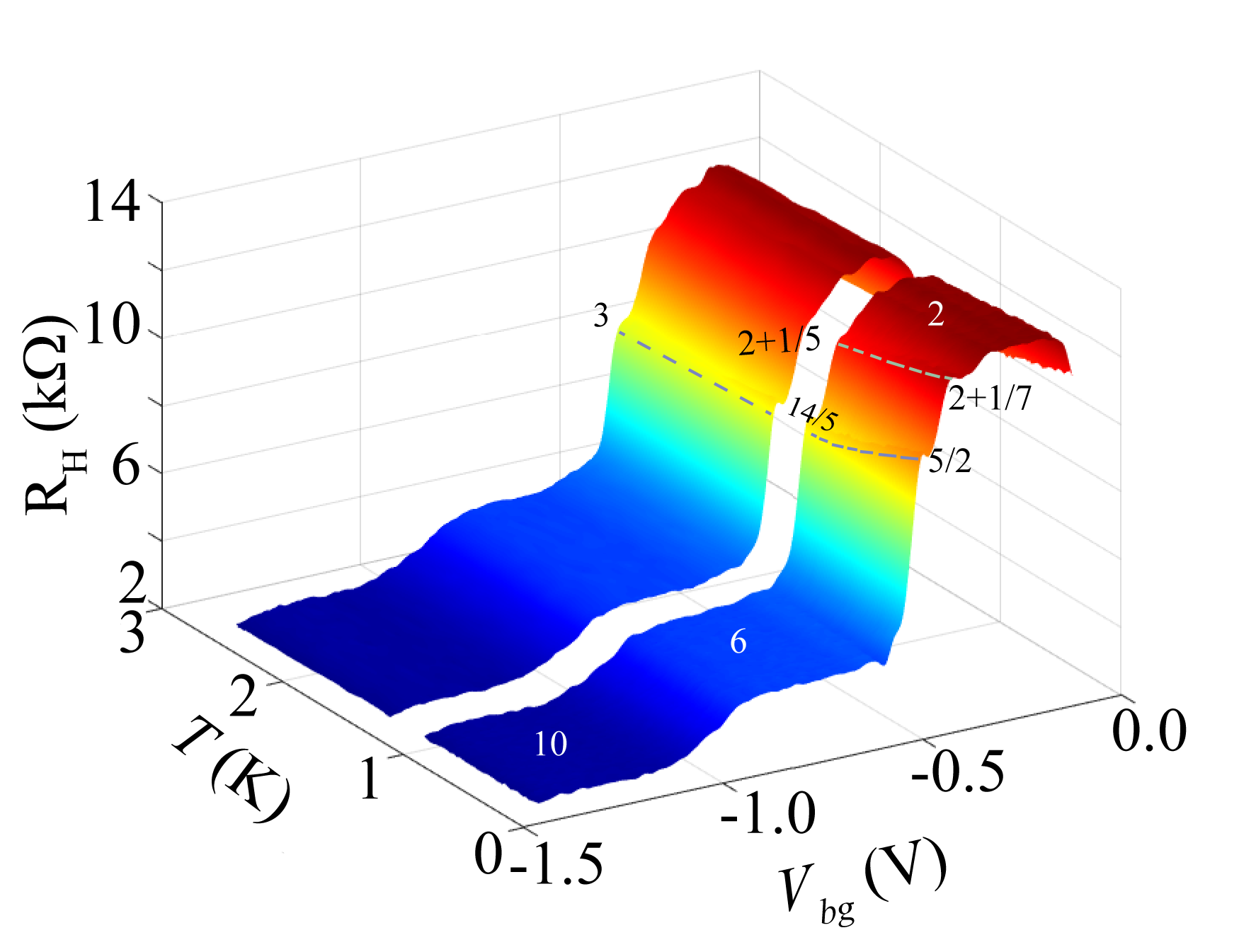}    
    \caption{The evolution of the plateau at $\nu_H = 5/2$ \textit{w.r.t.} temperature. The Hall resistance was measured at $B = -12.5$ T and the QPC is biased at -0.43 V. The Hall plateaus $\nu_H = 2, 6,$ and $10$ are robust with the increasing measured temperature up to $T = 2.5$ K. On the other hand, the plateau at $5/2$ is observed at $T = 40$ mK which gradually shifts to $2+4/5$ at $T = 1$ K and finally to $3$ at $T = 2.5$ K. Another plateau is seen at $2+1/7$ at $T = 40$ mK which evolves to $2+1/5$ at $T = 1$ K. }
    \label{FigTevolve_nu5by2}
\end{figure}

The standard Hall plateaus at $|\nu_H|= 2, 6 $ and 10 are robust \textit{w.r.t.} the top gate voltage $V_\text{tg}$ for several fixed magnetic fields at the electronic temperature of $T = 40$ mK. Here we present the study of $\nu_H = 5/2$ and integer state $\nu_H = 3$ for $B = -12.5$ T. The plateau at $\nu_H = 5/2$ evolves to $\nu_H = 3$ with increase in temperature. Also, the plateau at $\nu_H = 2+1/7$ evolves to $\nu_H = 2+1/5$, see \cref{FigTevolve_nu5by2}. The stabilization of the plateau $\nu_H = 5/2$ requires full equilibration of the state $\nu_{\text{bg}} = 2+2/5$ and $\nu_{\text{QPC}} = 2+1/3$ (``Ordinary"). The formation of $\nu_{\text{QPC}}$ is related to either the emergence of many localized states or a network of compressible stripes in the QPC region due to non-linear screening of the potential fluctuations. In the FQH regime, the gap energies are lower resulting in the formation of the local states within the QPC region. As we presented, the full equilibration between the bulk edge states $\nu_{\text{bg}}$ and these localized states $\nu_{\text{QPC}}$ results in a modification in the effective Hall filling factor $\nu_H$. With the temperature change, the local states within the QPC may favor certain filling fractions depending on their gap energy or even the degree of the coupling between bulk states and states in the QPC region. At much higher temperatures, the localized states within the confined region of QPC are completely washed out leading to effective $\nu_{\text{QPC}}$ = $\nu_{\text{bg}}$ = $\nu_H$. This is the case for the evolution of the $\nu_H = 5/2$ at $T = 40$ mK to $\nu_H = 3$ at $T = 2.5$ K.     

\subsection{Top gate dependence for regions I and II} \label{Note5b}
\begin{figure}[htb]
    \centering
    \includegraphics[width=0.9\textwidth]{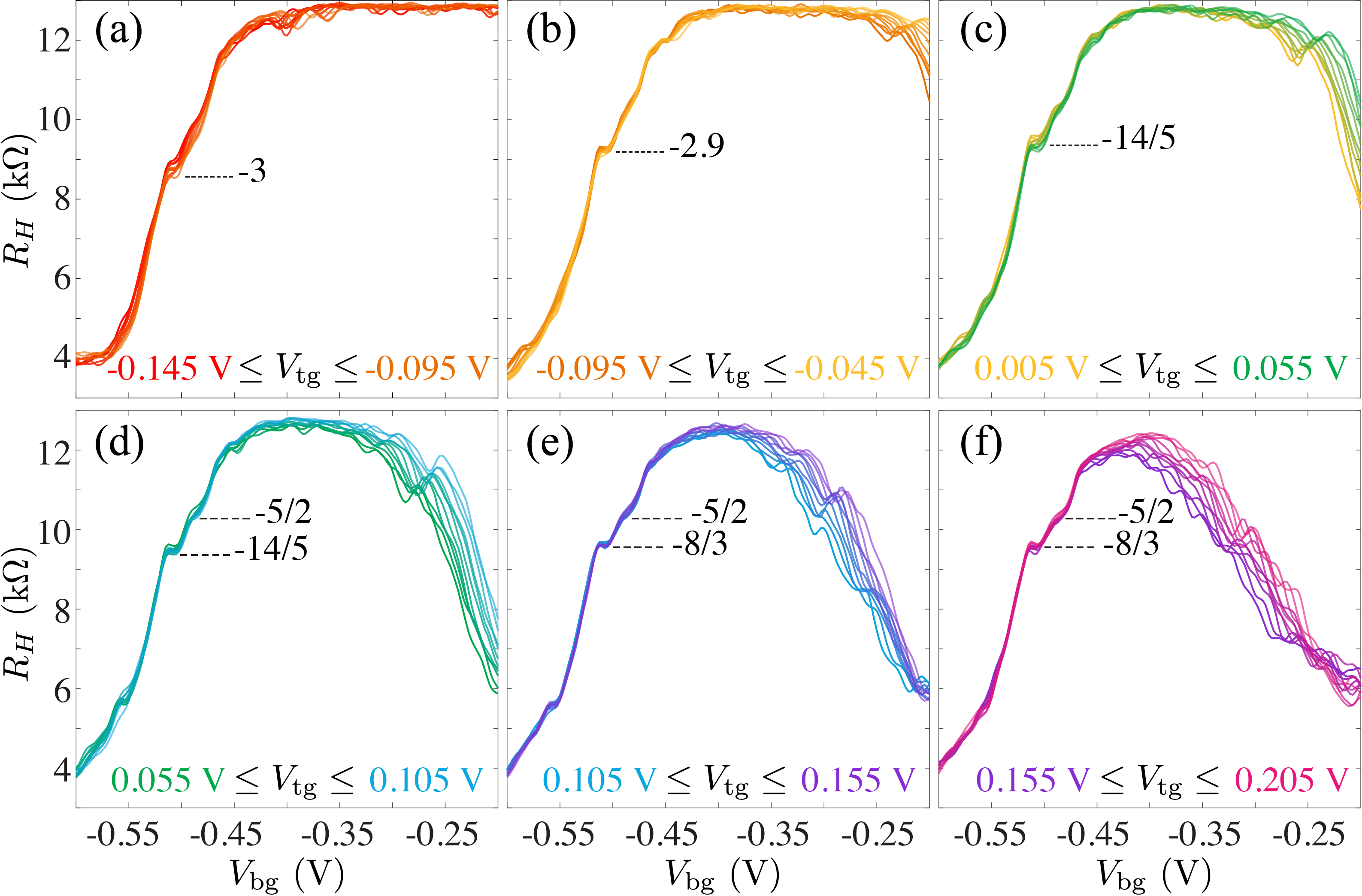}
    \caption{The evolution of the plateau at $\nu_H = -5/2$ \textit{w.r.t.} split top gate, region I: The Hall resistance was measured at $B = 10.75$ T and $T = 15$ mK. The top gate is sequentially changed from $V_\text{tg} = -0.145 V$ to $0.205$ V from (a) to (f). The corresponding top gate $V_\text{tg}$ is mentioned in each figure. Note that the filling fraction is negative. Compared to the fractional fillings, the Hall plateaus $\nu_H = -2, -6,$ and $-10$ are more robust \textit{w.r.t.} perturbation due to the change of the top gate potential. The plateau at $-5/2$ is observed as shoulder in top gate regime $ 0.055\leq V_\text{tg} \leq 0.205$. Another plateau corresponding to $\nu_H = -3$ is observed, which eventually evolves to -2.8 at $0.055\leq V_\text{tg} \leq 0.105$ and gradually shifts to $\nu_H = -2-2/3$ for $V_\text{tg} \geq 0.105$V.}
    \label{Gatedependence_1}
\end{figure}

\begin{figure}[htb]
    \centering
    \includegraphics[scale=1]{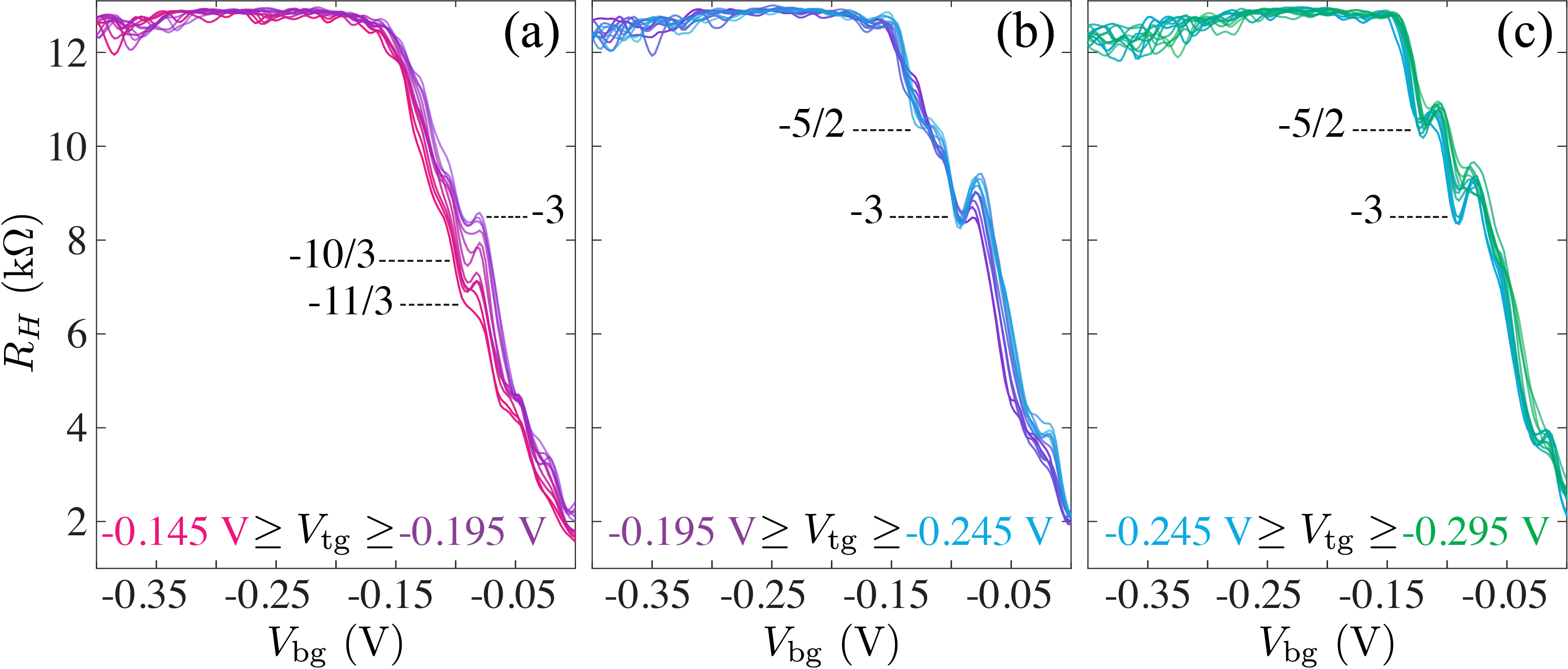} 
    \caption{The evolution of the plateau at $\nu_H = -5/2$ \textit{w.r.t.} split top gate, region III: The Hall resistance was measured at $B = 10.75$ T and $T = 15$ mK. The top gate is sequentially changed from $V_\text{tg} = -0.145 V$ to $-0.295$ V (a) to (c). Note that the filling fraction is negative. The plateau at $-5/2$ is observed as shoulder in top gate regime $-0.195 \text{ V } \geq V_\text{tg} \geq -0.245$ V and eventually developed as a dip-peak, \textit{i.e.} resonant feature. Another plateau corresponds to $\nu_H = -3$, which evolves from -10/3 and -11/3 starting at $V_\text{tg}= -0.145$ V to more negative top gate voltages.}
    \label{Gatedependence_2}
\end{figure}

For the gate dependence, the magnetic field was fixed at $B=10.75$ T and the temperature at $T =15$ mK. The top and back gates were swept between -2 V to 2 V. Here we present one set of such measurements with a snap shot of two gate voltage sweeps. We note that here the filling fraction is negative for regions I (\cref{Gatedependence_1}a-f) and II (\cref{Gatedependence_2}a-c) where $\nu_H = -5/2$ is observed. Similar instability of the plateau is observed \textit{w.r.t.} the magnetic field, see \cref{sevenby2evolve}b. 

\par Such instability in the fragile plateaus can be understood in terms of non-linear screening of the potential within the QPC region, which may result in the formation of locally enhanced or reduced density of states. The formation of these localized states is modulated by the gate potential, either by the back gate or the top gate. Similar effects may happen with a change in the magnetic field. The coupling between these localized states may stabilize more favorable local states in the QPC region. The effective coupling between the bulk state and local state supports a certain fraction (see \cref{Note6a}) \cite{Kumar2018}.  

\section{Stability of the plateau at the other half fillings}
\begin{figure}[htb!]
    \centering
    \includegraphics[scale=1]{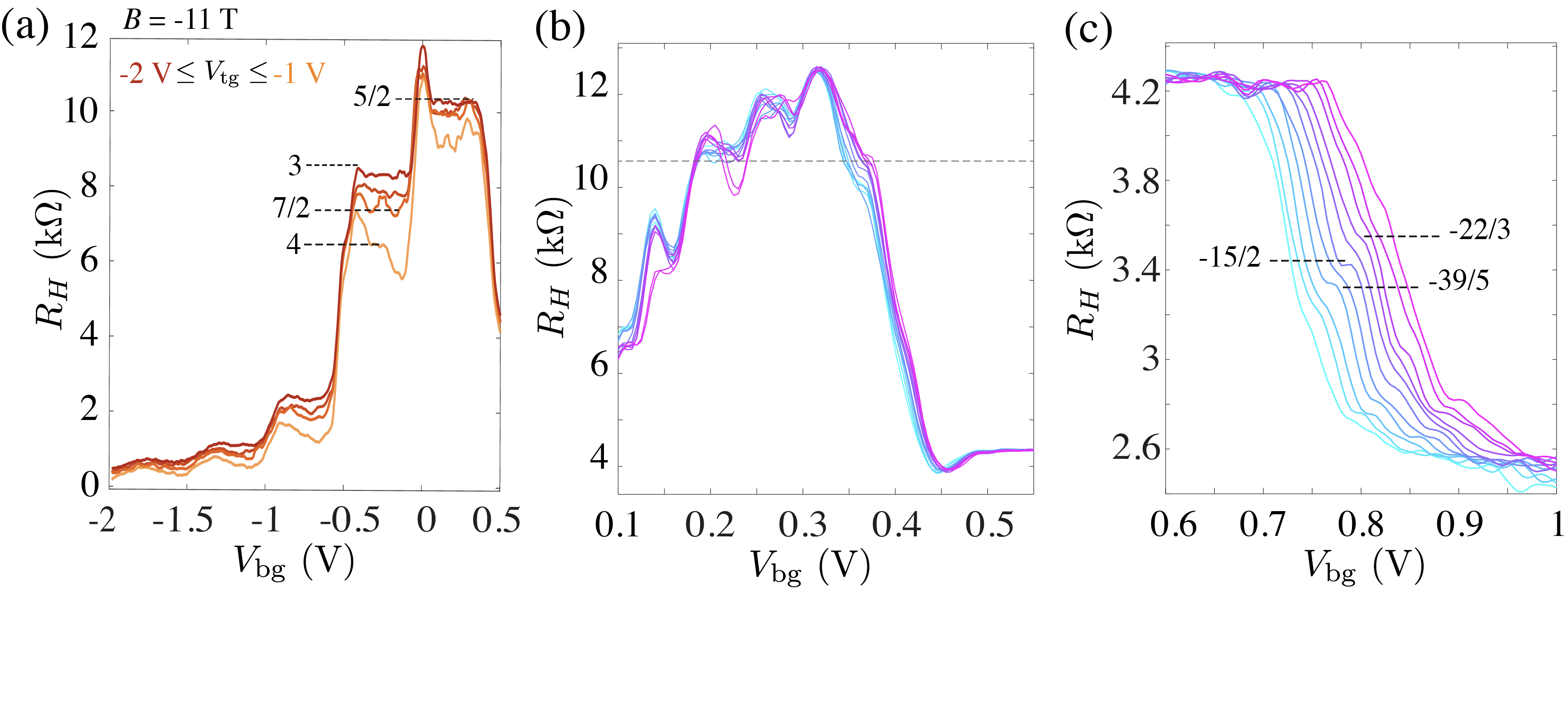}    
    \caption{The evolution of the plateau at $\nu_H = 7/2$, 5/2 and $15/2$ \textit{w.r.t.} gate and magnetic field. (a) The Hall resistance $R_H$ was measured at $B = -11$ T, and the back gate was varied with fixed voltage on the split gate, marked in the figure. The total transmission is heavily influenced by the edge states under the top gate. (b) The evolution of the $\nu_H = -5/2$ (region IV) from the $\nu_H = -2-2/3$ with the change of magnetic field from $-9.72$ T (magenta) to $-9.42$ T (cyan), with the traces measured in steps of $25$ mT. (c) Evolution of the quantize step at $\nu_H = -7-4/5$ to $\nu_H = -7-1/3$ with intermediate steps at $\nu_H = -7-1/2$. Here, traces are measured in steps of $125$ mT, starting from $B = -9.2$ T (cyan) to $B = -10.5$ T (end). The top gate for (b) and (c) is $V_\text{tg} = -0.75$ V. }
    \label{sevenby2evolve}
\end{figure}

 In our experiment, other quantization at half fillings is $\nu_H = 7/2, 9/2$, and $-15/2$. In \cref{sevenby2evolve}a, $R_H$ was measured with respect to the back gate voltage at four given top gate voltages for $B=-11$ T. For range $-0.5<V_\text{bg}<0$, the quantized steps evolve from $\nu_H = 3$ ($V_\text{tg}=-2 V$) to $10/3$, $3.47 \sim 7/2$ and $4$ ($V_\text{tg}=-1 V$); the steps at $-1.0<V_\text{bg}<-0.5$ evolve similarly from $\nu_H = 11$ to $12, 13$ and $\sim$ 15.33. The most striking are the plateaus for the back gate voltage range $0<V_\text{bg}<0.5$, with quantized steps at filling factor $\nu_H = 5/2$ at $V_\text{tg} = -2$ V, which evolves from $\nu_H = 14/5$ via $\nu_H = 8/3$. Note that here $\nu_H = 2$ is not fully developed. At the peak close to $V_\text{bg} = 0$ V, the filling factor hovers around $\nu_H = 7/3$. In our experiment, $\nu_H = 5/2$ is dominantly formed in the regime where $|\nu_{\text{bg}}| \lesssim 2$. In \cref{sevenby2evolve}c, we are showing the evolution of $\nu_H = -15/2$ from $\nu_H = -7-4/5$ with the increase in magnetic field, where a similar trend is seen with $\nu_H = \pm 5/2$ (\cref{sevenby2evolve}b).    

\section{Gap energy calculation } \label{Note8}
\begin{figure}[htbp]
    \centering
    \includegraphics[width=0.5\textwidth]{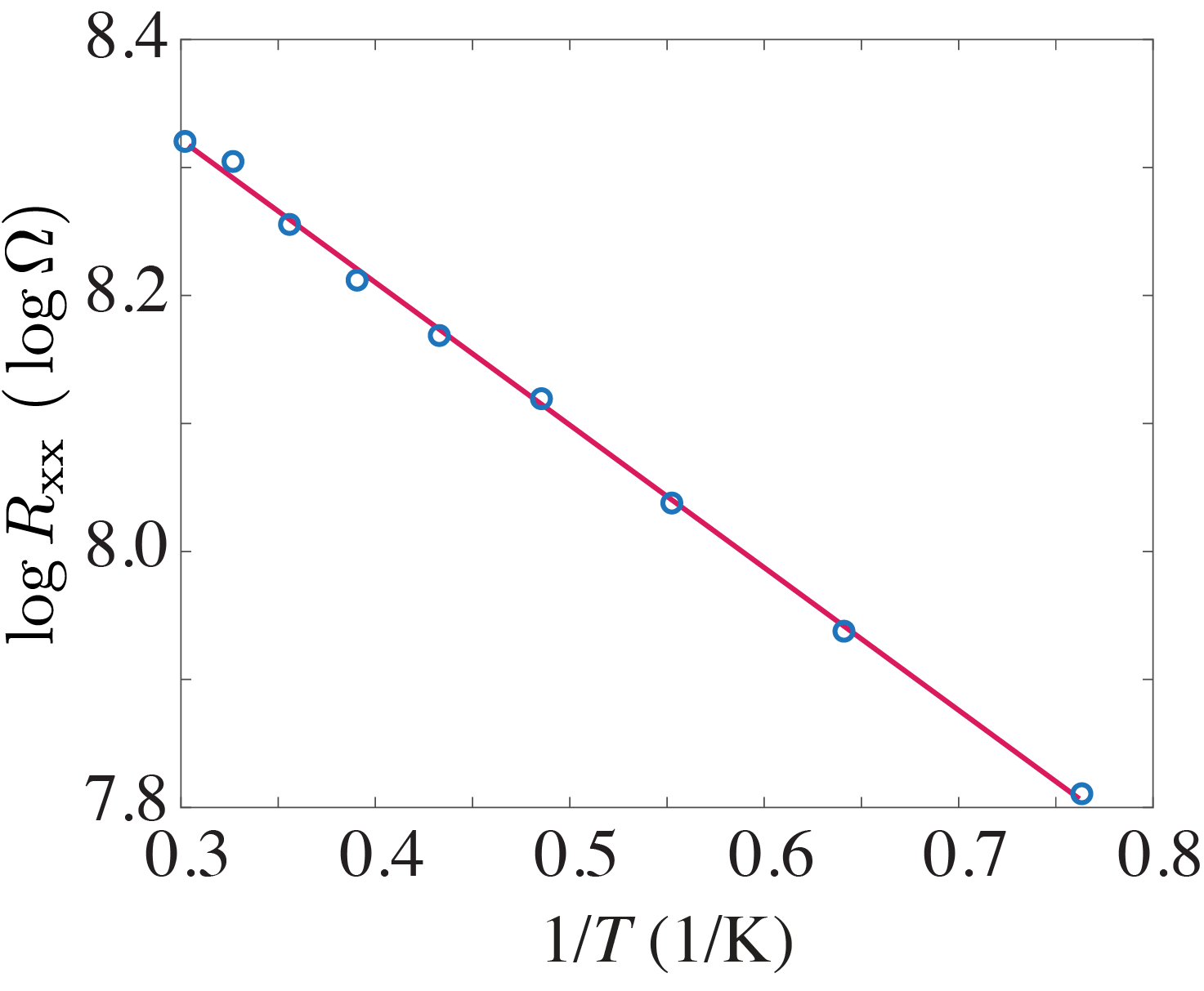}
    \caption{The Arrhenius activation gap energy calculation for $\nu_H = 2+1/3$ at $B = -10$ T. Here the top gate voltage $V_\text{tg} = 0$ V. Logarithmic of resistance $R_{xx}$ is plotted \textit{w.r.t.} inverse of the temperature. The linear fit (red) of the experimental data (blue) gives the estimation of the gap energy $\Delta = 1.14$ K  }
    \label{gapenergy_1}
\end{figure}

The dips in $R_{xx}$ for the corresponding plateaus in $R_{xy}$ are identified. To estimate the characteristic energy scale of the fractional quantum states, the temperature dependence of minimal values in $R_{xx}$ was studied.  The minima in $R_{xx}$ are expected to follow the thermally activated Arrhenius transport law $R_{xx} = R_0~\exp~(-T_0/T)$ where $T_0$ is a characteristic temperature of the system. First, we focus our studies on the standard FQH state $\nu_H = 2+1/3$. In \cref{gapenergy_1}, we plot the logarithm of resistance \textit{w.r.t.} the inverse of the temperature 1/$T$. The characteristic temperature $T_0 \simeq 1.14$ K represents the gap energy, \textit{i.e.} the quasiparticle-quasihole creation energy $\Delta = 2k_BT_0$, where $k_B$ is the Boltzmann constant. The obtained $T_0$ is quite small compared with typical transport gap values for 1/3 fractional states in graphene, a sign of disorder influencing the effective potential landscape in the present sample \cite{Ambrumenil2011}.

\par In higher Landau levels, the plateau at $R_H \sim 6.428$ k$\Omega$ ($\nu_H = 4$) is observed at lower $T \sim 68$ mK.  With the increase in the temperature, this plateau eventually evolves to $\nu_H = 3+1/5$ at $T = 1.2$ K with intermediate steps at $\nu_H = 3+1/2$ at $T = 0.7$ K. The Arrhenius activation fit for $0.6 \leq T \leq 1.2$ K yields $T_0 \simeq 185$ mK. The gap is thus strongly influenced by thermal fluctuations.
\par
\begin{figure}[ht!]
    \centering
    \includegraphics[width=\textwidth]{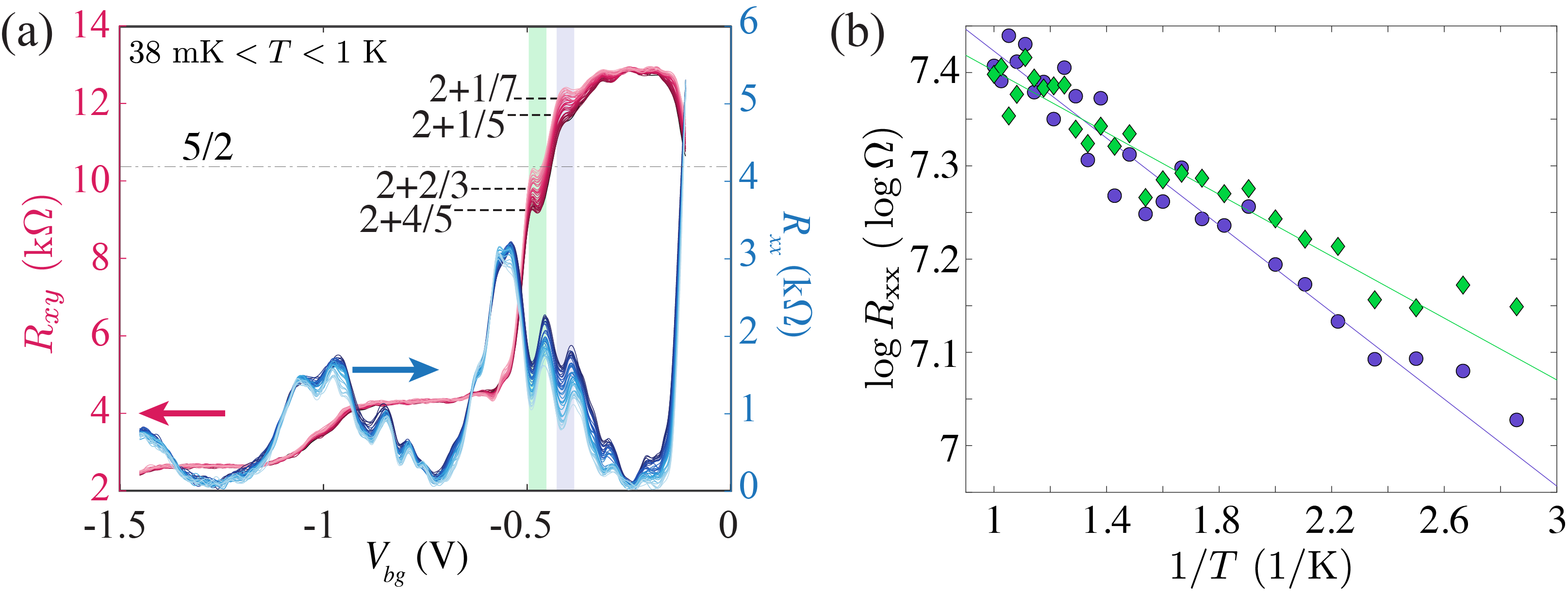}
    \caption{(a) The evolution of the plateau at 5/2 as a function of temperature from $T=1$ K (dark) to $T=38$ mK (light) at $B = -12.5$ T and $V_\text{tg} = 0.43$ V. $R_{xy}$ (magenta) and the corresponding $R_{xx}$ (blue) are shown for $V_\text{bg} < 0$. The plateau (marked by the purple-shaded region) is unstable and carries over to higher filling fractions, namely 2+2/3 and 2+4/5 as the temperature increases. The same applies to the plateau at 2+1/7, marked by the green-shaded region. The data presented here is part of \cref{FigTevolve_nu5by2}. (b) The corresponding thermal activation transport plot of (a) for the two marked regions, green ($\Diamond$): $2+1/2 \leq \nu_H \leq 2+4/5$ with $T_0 = 0.17$ K  and purple ($\circ$): $2+1/7 \leq \nu_H \leq 2+1/5$ with $T_0 = 0.23$ K. }
    \label{gapenergy_2}    
\end{figure}

 Contrary to the $\nu_H = 2+1/3$ state, the plateau at $\nu_H = 5/2$ in the region I shows the unusual shift to a higher filling factor with increasing temperature, while in the $R_{xx}$ measurement, the minima show lower dips as the temperature decreases (\cref{gapenergy_2}a).  The activated transport for the plateau at $\nu_H \simeq 5/2$ measured at $B = -10$ T gives gap energy of $\Delta = 110$ mK. The similar activation plot for the plateau at $5/2 \leq \nu_H \leq 12/5$ in the temperature regime of $0.04 \leq T \leq 1.2$ K and at $B = -12.5$ T gives gap energy of $\Delta =  229$ mK. 

\par Here, the reduced gap energy seen in the thermal activation plots (\cref{gapenergy_2}b) is due to the presence of disorder. This disorder creates puddles of compressible and incompressible states in the 2-DEG. The thermally driven tunneling between these puddles forms dissipative channels which reduce the gap energy. Thus the measured energy gap $\Delta$ is smaller compared to the intrinsic gap energy $\Delta_i$, with the amount quantified by the disorder broadening potential parameter $\Gamma: \Delta = \Delta_i -\Gamma$ \cite{Ambrumenil2011}

\section{Derivation of theoretical equations in the main text and description of plateaus under full charge equilibration} \label{Note6a}

\begin{figure}[htbp]
    \centering
        \includegraphics[width=0.95\textwidth]{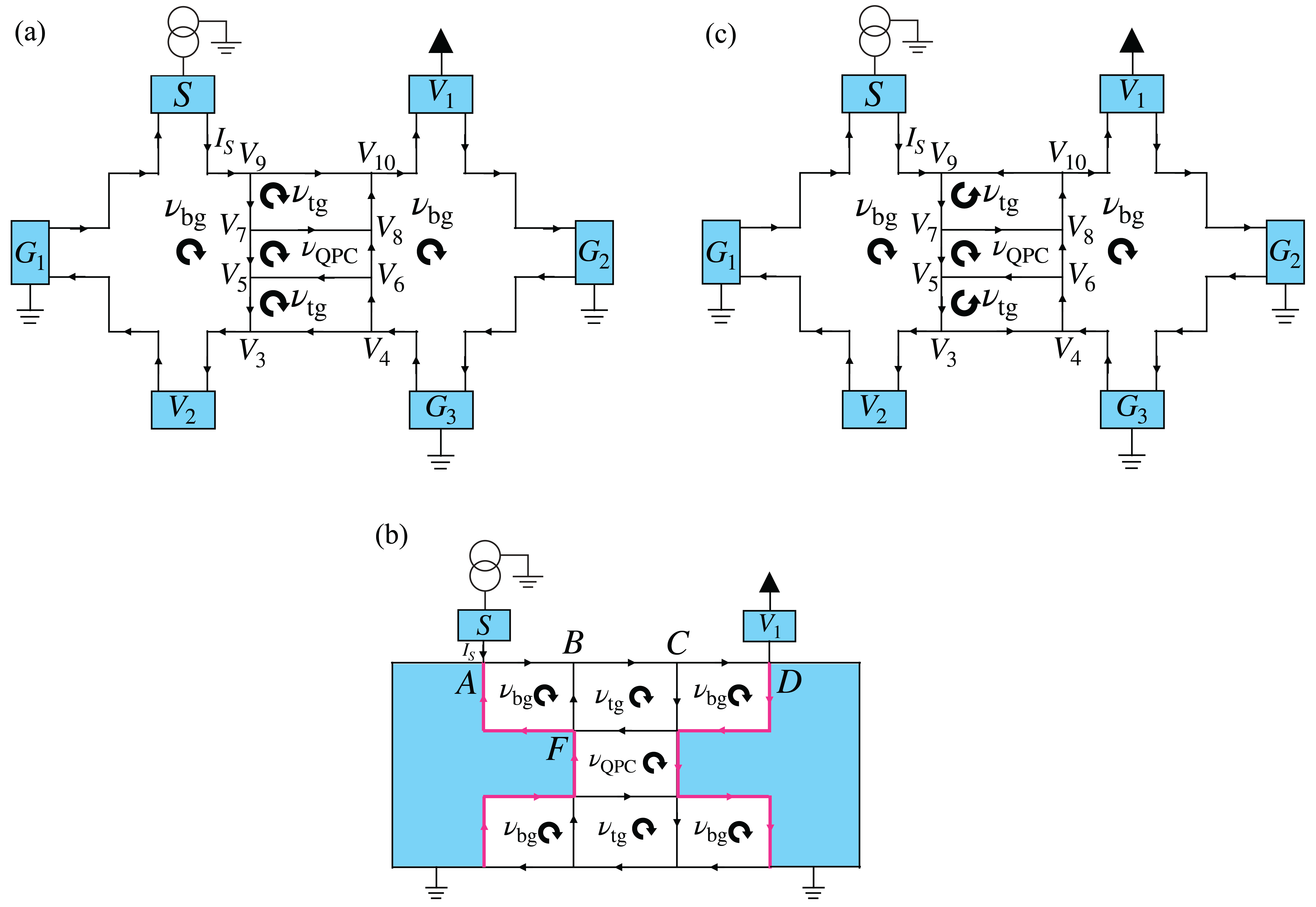}         \caption{Detailed pictures used in deriving the equations. We note that $V_1$ is equivalent to $V_H$ as in the main text. (a) Region I: $\nu_{\text{bg}}, \nu_{\text{tg}}, \nu_{\text{QPC}}$ are negative and $\nu_{\text{bg}}<\nu_{\text{QPC}}<\nu_{\text{tg}}$. (b) Region II ($V_\text{bg} < 0$) and III ($V_\text{bg} > 0$): 
        $ \nu_{\text{bg}}, \nu_{\text{QPC}}, \nu_\text{tg}$ are negative and $\nu_{\text{tg}}<\nu_{\text{bg}},\ \nu_{\text{tg}}<\nu_{\text{QPC}}$. (c) Region IV:
        $ \nu_{\text{bg}}, \nu_{\text{QPC}}$ are positive, $\nu_\text{tg}$ is negative and $\nu_{\text{tg}}<\nu_{\text{QPC}}<\nu_{\text{bg}}$. The pictures are also valid while changing the overall chirality of each filling and thereby adjusting the corresponding inequalities.}
\label{ManoharFig_3}
\end{figure}
We derive the theoretical equations in detail as shown in the main text and refer to \cref{ManoharFig_3}. We use those equations to explain the plateaus in \cref{Gatedependence_1}c and \cref{Gatedependence_2}a.

Region I $\text{(``Ordinary")}$: \cref{ManoharFig_3}a refers to the corresponding scenario.
We assume full charge equilibration at each segment of the device in the same spirit as illustrated in \cref{fig:cartoon}. 
This means that any geometric length ($L$) of our device is larger than the charge equilibration length $l^{\text{ch}}_{\text{eq}}$
leading to $L \gg l^{\text{ch}}_{\text{eq}}$.
This allows local voltages at each node of the setup to be defined. Then the next step is to write the Kirchhoffs current conservation law at each node and solve for the voltages. We explain in detail the corresponding equation for the node having local voltage $V_3$ as follows.
Here, incoming currents appear via two segments having the effective filling factors as $|\nu_{\text{tg}}|$ and $|\nu_{\text{bg}}|-|\nu_{\text{tg}}|$, respectively
carrying their voltages as $V_4$ and $V_5$. So, the total incoming current appearing at that node is  $|\nu_{\text{tg}}|V_4+(|\nu_{\text{bg}}|-|\nu_{\text{tg}}|)V_5$.
According to the Kirchhoffs current conservation law, this incoming current is equal to the outgoing current (flowing along the segment having filling $|\nu_{\text{bg}}|)$ from that node, which is $|\nu_{\text{bg}}|V_3$
resulting in an equation $|\nu_{\text{tg}}|V_4+(|\nu_{\text{bg}}|-|\nu_{\text{tg}}|)V_5=|\nu_{\text{bg}}|V_3$.
Similarly, as shown below, we can write all the equations at each node. We mention that we are using a current source injecting a current $I_S$ into the system. The corresponding voltage can be obtained using the respective filling fraction as shown below. The following equations at each node can be written by using charge equilibration:

\begin{equation}
    \begin{split}
        &V_{10}=V_1; V_3=V_2; |\nu_{\text{tg}}|V_4+(|\nu_{\text{bg}}|-|\nu_{\text{tg}}|)V_5=|\nu_{\text{bg}}|V_3; V_4=0;\\&(|\nu_{\text{bg}}|-|\nu_{\text{QPC}}|)V_7+(|\nu_{\text{QPC}}|-|\nu_{\text{tg}}|)V_6=(|\nu_{\text{bg}}|-|\nu_{\text{tg}}|)V_5; V_6=0; V_9=V_7;\\&(|\nu_{\text{QPC}}|-|\nu_{\text{tg}}|)V_7=(|\nu_{\text{bg}}|-|\nu_{\text{tg}}|)V_8; V_0=V_9; |\nu_{\text{tg}}| V_9+(|\nu_{\text{bg}}|-|\nu_{\text{tg}}|)V_8=|\nu_{\text{bg}}| V_{10}
    \end{split}
\end{equation}
where we have $V_0 = |\nu_{\text{bg}}|\frac{e^2}{h}/I_S$. Solving these, we find
\begin{equation}
    R_{H}=\frac{V_1}{I_S}=\frac{|\nu_{\text{QPC}}|}{|\nu_{\text{bg}}|^2} \frac{h}{e^2}.
\end{equation}

Region IV $\text{(``Ordinary")}$: Similarly, for the scenario described by \cref{ManoharFig_3}c, we write the following equations at each node:

\begin{equation}
    \begin{split}
        &V_{10}=V_1; V_3=V_2; V_5=V_3; |\nu_{\text{tg}}|V_3=(|\nu_{\text{bg}}|+|\nu_{\text{tg}}|)V_4;\\& (|\nu_{\text{bg}}|-|\nu_{\text{QPC}}|)V_7+(|\nu_{\text{tg}}|+|\nu_{\text{QPC}}|)V_6=(|\nu_{\text{bg}}|+|\nu_{\text{tg}}|)V_5; V_4=V_6; V_9=V_7;\\& (|\nu_{\text{tg}}|+|\nu_{\text{QPC}}|)V_7+(|\nu_{\text{bg}}|-|\nu_{\text{QPC}}|)V_6=(|\nu_{\text{tg}}|+|\nu_{\text{bg}}|)V_8;\\& |\nu_{\text{bg}}|V_0+|\nu_{\text{tg}}|V_{10}=(|\nu_{\text{bg}}|+|\nu_{\text{tg}}|)V_9; V_8=V_{10}
    \end{split}
\end{equation}
and find
\begin{equation}
    R_{H}=\frac{V_1}{I_S}=\frac{2|\nu_{\text{bg}}||\nu_{\text{tg}}|+|\nu_{\text{bg}}||\nu_{\text{QPC}}|-|\nu_{\text{tg}}||\nu_{\text{QPC}}|}{|\nu_{\text{bg}}|(|\nu_{\text{bg}}|^2+3|\nu_{\text{bg}}||\nu_{\text{tg}}|-2|\nu_{\text{tg}}||\nu_{\text{QPC}}|)} \frac{h}{e^2}.
\end{equation}

Region II and III $\text{(``Out-of-Ordinary")}$: Using \cref{ManoharFig_3}b we write:

\begin{equation}
    \begin{split}
        &I_S=V_0/2\rho+|\nu_{\text{bg}}|V_0; (|\nu_{\text{tg}}|+|\nu_{\text{bg}}|)V_0/2=|\nu_{\text{tg}}|V_B;\\& V_B=V_C; |\nu_{\text{bg}}|V_C=V_1/2\rho; (|\nu_{\text{tg}}|-|\nu_{\text{QPC}}|)V_1=|\nu_{\text{tg}}|V_0,
    \end{split}
\end{equation}
where $\rho$ is the resistance of the metal. Then we find
\begin{equation}
    R_{H}=\frac{V_1}{I_S}=\frac{2|\nu_{\text{tg}}|^3}{|\nu_{\text{bg}}|(|\nu_{\text{tg}}|-|\nu_{\text{QPC}}|)(|\nu_{\text{tg}}|(|\nu_{\text{bg}}|+3|\nu_{\text{tg}}|)-(|\nu_{\text{bg}}|+|\nu_{\text{tg}}|)|\nu_{\text{QPC}}|)} \frac{h}{e^2}.
\end{equation}
We theoretically explain the plateaus observed in \cref{Gatedependence_1}c and \cref{Gatedependence_2}a by using our calculations and show the results in \cref{Table2}.

\begin{table}[htbp]
\centering
\begin{tabular}{||c| c| c| c||} 
 \hline
 $|\nu_{H}|$ & $|\nu_{\text{tg}}|$ & $\{|\nu_{\text{bg}}|, |\nu_{\text{QPC}}|\}|$  & $\text{Region}$ \\ [0.5ex] 
 \hline\hline
 2+1/7 & 2 & \{2+1/9, 2+1/11\} & \multirow{16}{*}{I} \\
   \cline{1-3}
 2+1/5 & 2 & \{2+1/7, 2+1/9\} & \\ 
  \cline{1-3}
  2+1/3 & 2 & \{2+1/5, 2+1/9\} & \\ 
  \cline{1-3}
  2+1/2  & 2 & \{2+2/5, 2+1/3\} & \\ 
  \cline{1-3}
 2+2/3 & 2 & \{2+2/3, 2+2/3\} & \\
  \cline{1-3}
   2+4/5 & 2 & \{2+4/5, 2+4/5\} & \\
  \cline{1-3}
 3 & 3 & \{3, 3\} & \\
   \cline{1-3}
 3+1/5 & 3 & \{3+1/7, 3+1/9\} & \\
  \cline{1-3}
  3+1/3 & 3 & \{3+1/5, 3+1/9\} & \\
  \cline{1-3}
  3+1/2 & 3 & \{3+2/5, 3+1/3\} & \\
  \cline{1-3}
  3+4/5 & 3 & \{3+4/5, 3+4/5\} & \\
  \cline{1-3}
  4 & 4 & \{4, 4\} & \\
  \cline{1-3}
 6 & 6 & \{6, 6\} & \\
  \cline{1-3}
  6+1/5& 6 & \{6+1/7, 6+1/9\} & \\
  \cline{1-3}
  6+4/5& 6 & \{6+4/5, 6+4/5\} & \\
  \cline{1-3}
   7+1/3 & 7 & \{7+1/5, 7+1/9\} & \\ 
 \hline
  \hline
  2+1/7& 2 & \{1+4/5, 3/5\} & \multirow{6}{*}{II} \\
  \cline{1-3}
 2+1/2 & 2 & \{1+2/3, 1/3\} & \\ 
  \cline{1-3}
  2+2/3 & 2 & \{1+2/3, 2/9\} & \\ 
  \cline{1-3}
  3  & 3 & \{2+4/5, 1\} & \\ 
  \cline{1-3}
  3+1/3 & 3 & \{2+2/3, 4/5\} & \\ 
  \cline{1-3}
  3+2/3 & 3 & \{2+2/3, 2/3\} & \\ 
  \hline
   \hline
  2+1/3 & 2 & \{1+2/3, 2/5\} & \multirow{4}{*}{III} \\ 
  \cline{1-3}
  2+2/5 & 2 & \{1+2/3, 4/11\} &  \\  
  \cline{1-3}
   2+1/2 & 2 & \{1+2/3, 1/3\} & \\ 
  \cline{1-3}
    2+3/5 & 2 & \{1+3/5, 1/5\} & \\ 
  \hline
   \hline
  2+1/7 & 2 & \{2+1/9, 2+1/11\} & \multirow{7}{*}{IV} \\ 
  \cline{1-3}
  2+1/2 & 2 & \{2+2/5, 2+1/3\} & \\
  \cline{1-3}
  6+1/2 & 6 & \{6+2/5, 6+1/3\} & \\ 
  \cline{1-3}
  6+2/3 & 6 & \{6+2/3, 6+2/3\} & \\ 
  \cline{1-3}
 10+1/3 & 10 & \{10+1/5, 10+1/9\} & \\ 
  \cline{1-3}
  11 & 11 & \{11, 11\} & \\ 
  \cline{1-3}
 11+1/7 & 11 & \{11+1/9, 11+1/11\} & \\ 
  \hline
\end{tabular}

\caption{Fractions close to calculated values for quantized conductances $\nu_H$ with their expected back gate and QPC filling fractions $\{|\nu_{\text{bg}}|, |\nu_\text{QPC}| \}$ for given top gate filling factors in the range $|\nu_{\text{tg}}|=2-11$.}
\label{Table2}
\end{table}

\printbibliography	
\end{document}